\documentclass[usenatbib, english]{mnras}
\usepackage[fleqn]{amsmath}
\usepackage{array,multirow,graphicx}
\usepackage{txfonts}
\usepackage{lmodern}
\bibpunct{(}{)}{;}{a}{}{,}
\usepackage{floatrow}
\usepackage{array}
\usepackage{relsize}
\usepackage[font=small,labelfont=bf]{caption}
\usepackage{bm}

\usepackage{listings}

\newfloatcommand{capbtabbox}{table}[][\FBwidth]

\def\app#1#2{%
\mathrel{%
\setbox0=\hbox{$#1\sim$}%
\setbox2=\hbox{%
\rlap{\hbox{$#1\propto$}}%
\lower1.1\ht0\box0%
}%
\raise0.25\ht2\box2%
}%
}
\def\approxprop{\mathpalette\app\relax}

\title[How to get cool in the heat]{How to get cool in the heat: comparing analytic models of hot, cold, and cooling gas in haloes and galaxies with EAGLE}
\author[A.~R.~H.~Stevens et al.]{Adam R.~H.~Stevens,$^1$\thanks{E-mail: astevens@astro.swin.edu.au} 
Claudia del P.~Lagos,$^{2,3}$ 
Sergio Contreras,$^4$ 
Darren J.~Croton,$^1$ 
\newauthor Nelson D.~Padilla,$^{4,5}$ 
Matthieu Schaller,$^6$ 
Joop Schaye$^7$ 
and Tom Theuns$^6$\\
$^1$Centre for Astrophysics \& Supercomputing, Swinburne University of Technology, Hawthorn, VIC 3122, Australia\\
$^2$International Centre for Radio Astronomy Research, The University of Western Australia, Crawley, WA 6009, Australia\\
$^3$Australian Research Council Centre of Excellence for All-sky Astrophysics (CAASTRO), Redfern, NSW 2016, Australia\\
$^4$Instituto de Astrof\'{i}sica, Pontificia Universidad Cat\'{o}lica de Chile, Santiago, Chile\\
$^5$Centro de Astro-Ingenier\'{i}a, Pontificia Universidad Cat\'{o}lica de Chile, Santiago, Chile\\
$^6$Institute for Computational Cosmology, Durham University, Durham, DH1 3LE, United Kingdom\\
$^7$Leiden Observatory, Leiden University, 2300 RA Leiden, The Netherlands}

\begin{document}

\pagerange{\pageref{firstpage}--\pageref{lastpage}} \pubyear{2016}

\maketitle

\label{firstpage}

\begin{abstract}
We use the hydrodynamic, cosmological EAGLE simulations to investigate how hot gas in haloes condenses to form and grow galaxies.  We select haloes from the simulations that are actively cooling and study the temperature, distribution, and metallicity of their hot, cold, and transitioning `cooling' gas, placing these in context of semi-analytic models.  Our selection criteria lead us to focus on Milky Way-like haloes.  We find the hot-gas density profiles of the haloes form a progressively stronger core over time, the nature of which can be captured by a $\beta$ profile that has a simple dependence on redshift.  In contrast, the hot gas that will cool over a time-step is broadly consistent with a singular isothermal sphere. We find that cooling gas carries a few times the specific angular momentum of the halo and is offset in spin direction from the rest of the hot gas. The gas loses $\sim$60\% of its specific angular momentum during the cooling process, generally remaining greater than that of the halo, and it precesses to become aligned with the cold gas already in the disc.  We find tentative evidence that angular-momentum losses are slightly larger when gas cools onto dispersion-supported galaxies.  We show that an exponential surface density profile for gas arriving on a disc remains a reasonable approximation, but a cusp containing $\sim$20\% of the mass is always present, and disc scale radii are larger than predicted by a vanilla \citeauthor{fall80} model.  These scale radii are still closely correlated with the halo spin parameter, for which we suggest an updated prescription for galaxy formation models.
\end{abstract}


\section{Introduction}
\label{sec:intro}
Fundamentally, galaxies must form from the cooling and condensation of gas residing in overdense regions of the Universe \citep[e.g.][]{white78}.  The manner in which the gas is accreted over cosmic time is vital to the structure of a galaxy and how it evolves.  This is especially true for late-type galaxies, where discs hold the majority of a galaxy's baryonic mass, and when considering the high-redshift Universe, prior to mergers dominating the growth of the most massive galaxies.  

The semi-analytic approach of galaxy formation (\citealt*{white91,kauffmann93,cole94,kauffmann99}; see reviews by \citealt{baugh06,somerville15}) provides a framework that is consistent with the hierarchical assembly of overdense regions, known as haloes.  Here, the histories and properties of dark-matter haloes formed in a cosmological $N$-body simulation provide the input for the coupled differential equations that describe the evolution of galaxies.  These computationally efficient models have proven highly successful in their ability to reproduce the statistical properties of galaxies in the local Universe and are becoming progressively more successful in the early Universe \citep[e.g.][]{gp14,henriques15,lacey16}.

Many of the semi-analytic models in the literature \citep*[e.g.][]{cole00,hatton03,croton06,sage,lagos08,somerville08,guo11,benson12} have been, in part, based on the disc formation scenario developed by \citet{fall80} and \citet*{mo98}.  In these models, it is assumed that halo gas (the circumgalactic or intracluster medium) has a uniform temperature and carries the same total specific angular momentum as the halo \citep[variants have adopted a statistical take on spin direction -- see][]{padilla14,lagos15a}.  Two regimes of gas accretion onto galaxies are typically employed.  For the `hot mode', infalling gas is first shock heated to the virial temperature, then must cool to reach the galaxy.  In the `cold mode', the cooling time-scale is small relative to the free-fall time-scale, and hence the former no longer sets the time-scale for accretion \citep[see][]{white91,birnboim03,keres05,benson11}.  When the gas condenses onto a disc, it is assumed to conserve its specific angular momentum, and to settle with a surface density that decreases exponentially with radius.  Often rotation curves of discs are approximated as completely flat.  Variants also include exponential cooling profiles as a function of specific angular momentum \citep*{stevens16}.  While these assumptions all carry physical merit, like anything, they would ideally be subject to independent testing.

With this study, we are motivated to investigate the basic assumptions of the aforementioned models with respect to the hot mode of accretion.  That is, how is hot gas distributed in haloes, what is the nature and distribution of gas as it cools onto a galaxy, and does this gas conserve its specific angular momentum?  These are not only important science questions in and of themselves, but answers to these have a direct impact on galaxy evolution model development.  To address these topics with modern observations would be an unsurmountable challenge.  In this sense, we cannot \emph{truly} test the model assumptions.  We can, however, look to numerical experiments with more detailed physics for insight, namely hydrodynamic cosmological simulations.  This allows us to see how the widely adopted theoretical description of gas cooling on global scales \citep{fall80,white91} compares to predictions from modelling galaxy evolution physics on local ($\lesssim$ kpc) scales.  Hydrodynamic simulations also carry the advantage that we can immediately relate any results regarding gas cooling to the properties of dark-matter haloes.

We use the EAGLE simulations \citep[Evolution and Assembly of GaLaxies and their Environments;][]{schaye15,crain15} to study the gas particles in haloes that cool from a hot state down onto galaxies.  We assess how these particles are distributed in physical space and in terms of specific angular momentum both prior to and after cooling episodes.  We compare this against the overall hot and cold gas profiles in haloes, investigating how the process of cooling leads to evolution in these structures.  We further determine whether the angular momentum of these particles is conserved as they cool, both on an individual basis and as a collective system.  EAGLE provides an ideal testbed to address these questions, as there is growing evidence the simulation produces a realistic galaxy population in terms of mass \citep{furlong15}, size \citep{furlong16}, specific angular momentum \citep{lagos16b}, and gas content \citep{lagos15b,bahe16}.
 
Recently, \citet{guo16} compared the galaxy populations of two semi-analytic models, run on the dark-matter-only halo merger trees of EAGLE, against the main hydrodynamic EAGLE simulation.  Those authors found consistency in the evolution of the stellar mass function and the specific star formation rates of galaxies, but noted clear differences when galaxies were broken into passive and star-forming.  However, their study did not address whether the physical description and approximations of galaxy evolution processes in the semi-analytic models were supported by hydrodynamic simulations.  Our study contributes by focussing on the physics of the models, rather than testing whether the end result, i.e.~the galaxy properties, are in agreement with each other or observations. 
 
While we address aspects of models of halo gas cooling in this paper, our focus has intentionally not been placed on raw cooling rates.  Comparisons of cooling rates in hydrodynamic simulations and semi-analytic models have already been the focus of papers in the past \citep{lu11,monaco14}.  A major challenge for these types of studies is the complex interplay between cooling and feedback, and the differing techniques for implementing this in hydrodynamic simulations and semi-analytic models.  For example, a semi-analytic model must \emph{explicitly} distinguish between feedback that reheats cold gas and feedback that suppresses cooling of hot gas \citep[see, e.g.,][]{croton06}, which might instead be \emph{implicit} in a hydrodynamic simulation.  Furthermore, the knowledge that gas has been reheated is typically not maintained in a semi-analytic model, yet hydrodynamic simulations have shown that the same gas particles can be reheated and cool onto the same galaxy many times (\citealt{christensen16}; also see \citealt{oppenheimer10,ford14}).  This adds challenges related to time resolution for extracting the true cooling rates of haloes from the snapshot data of hydrodynamic simulations.  As a result of all this, feedback has been favourably omitted for the aforementioned comparison studies, which hence assess idealised, rather than physical, cooling rates.  In light of these challenges, but with a different method of approach, we compare semi-analytic-equivalent cooling rates of EAGLE haloes using their full structural information from the simulations versus the standard method of using their global properties.  A complete study of the physical cooling rates of EAGLE haloes is worthy of a paper by itself, and thus we leave that for future work.
 
This paper is structured as follows.  In Section \ref{sec:eagle}, we provide details on the design of the EAGLE simulations and the data products that we have used.  We also specify the sample of galaxies within the simulations that we study.  In Section \ref{sec:hot}, we examine the state of hot gas in haloes, comparing this to the particles which are about to cool.  In Section \ref{sec:angmom}, we discuss the direction and magnitude, with respect to the halo, of the specific angular momentum of gas as it cools, and whether this is conserved.  Once those particles have cooled onto a disc, we study their distribution, and that of the disc in full, in Section \ref{sec:cold}.  The results of this paper are then summarised in Section \ref{sec:summary}.

Throughout this paper, where relevant, we assume the cosmology used in the EAGLE simulations (see Section \ref{sec:eagle}).  We also use the term `virial radius' as the radius enclosing an average density 200 times greater than the critical density of the Universe.  I.e.~$R_{\rm vir} = R_{200}$, $M_{\rm vir} = M_{200} = M(<R_{200})$.  We use capital $R$ for three-dimensional radial distances ($R^2 = x^2 + y^2 + z^2$) and lower-case $r$ for two-dimensional equivalents ($r^2 = x^2 + y^2$, where the $z$-direction is always parallel to the relevant rotation axis).

\section{The EAGLE simulations and data}
\label{sec:eagle}
\subsection{Overview of the simulations}

The EAGLE simulations are a suite of state-of-the-art hydrodynamic, cosmological simulations, first presented by \citet{schaye15}.  They were run using a significantly modified version of \textsc{gadget3}, an $N$-body Tree-PM smoothed-particle hydrodynamics (SPH) code.  Various modifications of \textsc{gadget} have been developed by the community.  The most recent official public release was \textsc{gadget2} \citep{springel05}.  The simulations assumed a $\Lambda$CDM cosmology with parameters based on the 2013 \emph{Planck} data release: $\Omega_M = 0.307$, $\Omega_\Lambda = 0.693$, $\Omega_b = 0.048$, $h=0.6777$, $\sigma_8 = 0.8288$, and $n_s = 0.9611$ \citep{planckXVI}.  The main `Reference' simulation used a dark-matter particle mass of $9.70 \times 10^6\,{\rm M}_{\odot}$, an initial gas particle mass of $1.81 \times 10^6\,{\rm M}_{\odot}$, and a Plummer-equivalent gravitational softening scale of 2.66 comoving kpc for $z>2.8$ and 700 physical pc otherwise.  The simulations implemented a modified version of SPH, dubbed \textsc{anarchy} (Dalla Vecchia in preparation; but see appendix A of \citealt{schaye15}), along with the time-step limiter of \citet{durier12}.  \citet{schaller15} presented a comparison of the simulations with the old SPH version, and found that the modifications were important for active galactic nucleus (AGN) activity and star formation rates of large galaxies, but, by and large, did not have a big impact on stellar masses and sizes.

The simulations include a set of sub-grid models that describe physical processes below the resolution limit.  These are described in full in \citet{schaye15}.  Briefly, these include the following.
\begin{itemize}
\item Radiative cooling of gas particles is calculated by following 11 elements \citep*{wiersma09a}, which are assumed to be in ionization equilibrium. Gas is exposed to the cosmic microwave background, and after $z=11.5$, a uniform ionizing background is included \citep{haardt01}.  Self-shielding and local stellar radiation are ignored.
\item Star formation is based on the local pressure of gas \citep{schaye08} and is consistent with the Kennicutt-Schmidt relation \citep{kennicutt98}.  This is only triggered above a threshold local density, which has a metallicity-dependent value \citep{schaye04}.  Star particles are each considered to be single stellar populations that follow the initial mass function of \citet{chabrier03} and the stellar evolution models described by \citet{wiersma09b}.  Stellar feedback from supernovae is implemented with the stochastic thermal model of \citet{dalla12}.
\item Black-hole particles of mass $10^5h^{-1}\,{\rm M}_{\odot}$ are seeded in haloes once they reach a mass of $10^{10}h^{-1}\,{\rm M}_{\odot}$ \citep*{springel05b}.  These accrete gas from neighbouring particles, with consideration of angular momentum as in \citet{rosas15}, which leads to AGN feedback where nearby gas particles are stochastically heated \citep{schaye15}.
\end{itemize}
The free parameters of the sub-grid physics in EAGLE were calibrated to best reproduce a small number of key observables of galaxies in the local ($z \sim 0$) Universe: the stellar mass function \citep{li09,baldry12}, the stellar size--mass relation of \citet{shen03}, and the black hole--stellar mass relation of \citet{mcconnell13}.  See \citet{crain15} for further details.

In this paper, in addition to the Reference simulation, we also use runs with modified feedback physics but equal resolution.  These include runs with stellar feedback half the Reference strength (WeakFB), stellar feedback with twice the Reference strength (StrongFB), and an absence of active galactic nucleus feedback (NoAGN).  We also use the high-resolution run of EAGLE (with 8-fold superior mass resolution), for which the strength of feedback was recalibrated to meet the same observational constraints as mentioned above (RecalHR).  The alternate-feedback and high-resolution runs were performed in 25-Mpc boxes.  As we want to fairly compare these to Reference haloes, we use the 25-Mpc Reference simulation for this paper as well.  Further details on these simulations can also be found in \citet{crain15}.

\subsection{Halo finding}
To identify structures in EAGLE, \textsc{subfind} \citep{subfind,dolag09} was run on the output of the simulations.  For the purposes of the code, a `halo' is a collection of particles found with a friends-of-friends algorithm.  A `subhalo' is a set of particles in a halo which is enclosed by isodensity contours that traverse saddle points, where only particles gravitationally bound to the substructure are included.  A subhalo encompasses a galaxy and its immediate surroundings.  All haloes have a minimum of one subhalo (haloes with more than one subhalo have satellite galaxies).  Neither haloes nor subhaloes are inherently restricted to their virial radii (we specify when only particles within $R_{\rm vir}$ are considered).  Note that all results in this paper concerning `haloes' only consider the primary subhalo.

\subsection{Galaxy sample selection}
\label{ssec:sample}
For the purposes of this paper, we are interested in studying haloes undergoing cooling episodes.  To identify a relevant sample of systems in EAGLE, first we select central subhaloes (most massive in their parent halo) with total gas masses $>10^{9.5}\,{\rm M}_{\odot}$, then we compare the state of gas particles of the same IDs at temporally adjacent snapshots within each subhalo.  More specifically, we take each relevant subhalo at snapshot number $s$ and tabulate the IDs of all particles we consider `hot'.  We then find the same particles at snapshot $s+1$ and determine which of these are now `cold' and within the same system.  Those that have transitioned from hot to cold, or from hot to a star particle, from snapshot $s$ to $s+1$ are labelled `cooled' particles.  Equivalently, we refer to these as `cooling' particles at snapshot $s$.  If the number of cooling particles exceeds a threshold value (64, see below), then we deem a resolved cooling episode to have happened and include that subhalo in our sample.  We perform this same process for all snapshot pairs.  

Note that in order to have a clean sample of central galaxies to compare against how semi-analytic models treat centrals, we exclude galaxies from our analysis that are satellites at either snapshot (that is, we only consider the primary subhalo of each halo).  We further exclude galaxies that undergo a major merger between snapshots. We define a major merger as one where the smaller galaxy has a stellar mass at least 30\% that of the central.  These would have only constituted a small fraction of our sample.  As such, had we included them, our results would not have noticeably changed.  When selecting haloes from the alternate-feedback and high-resolution runs, we add a requirement that only haloes also present in the Reference sample can be included.

Throughout this text, we define `cold' particles as either having non-zero star formation rates or having both temperature $T<10^{4.5}\,{\rm K}$ and hydrogen number density $n_{\rm H} > 0.01\,{\rm cm}^{-3}$, which should capture the interstellar medium of most galaxies.  Conversely, `hot' gas particles have zero star formation rates and $T>10^5\,{\rm K}$.

\begin{figure}
	\centering
	\includegraphics[width=0.95\textwidth]{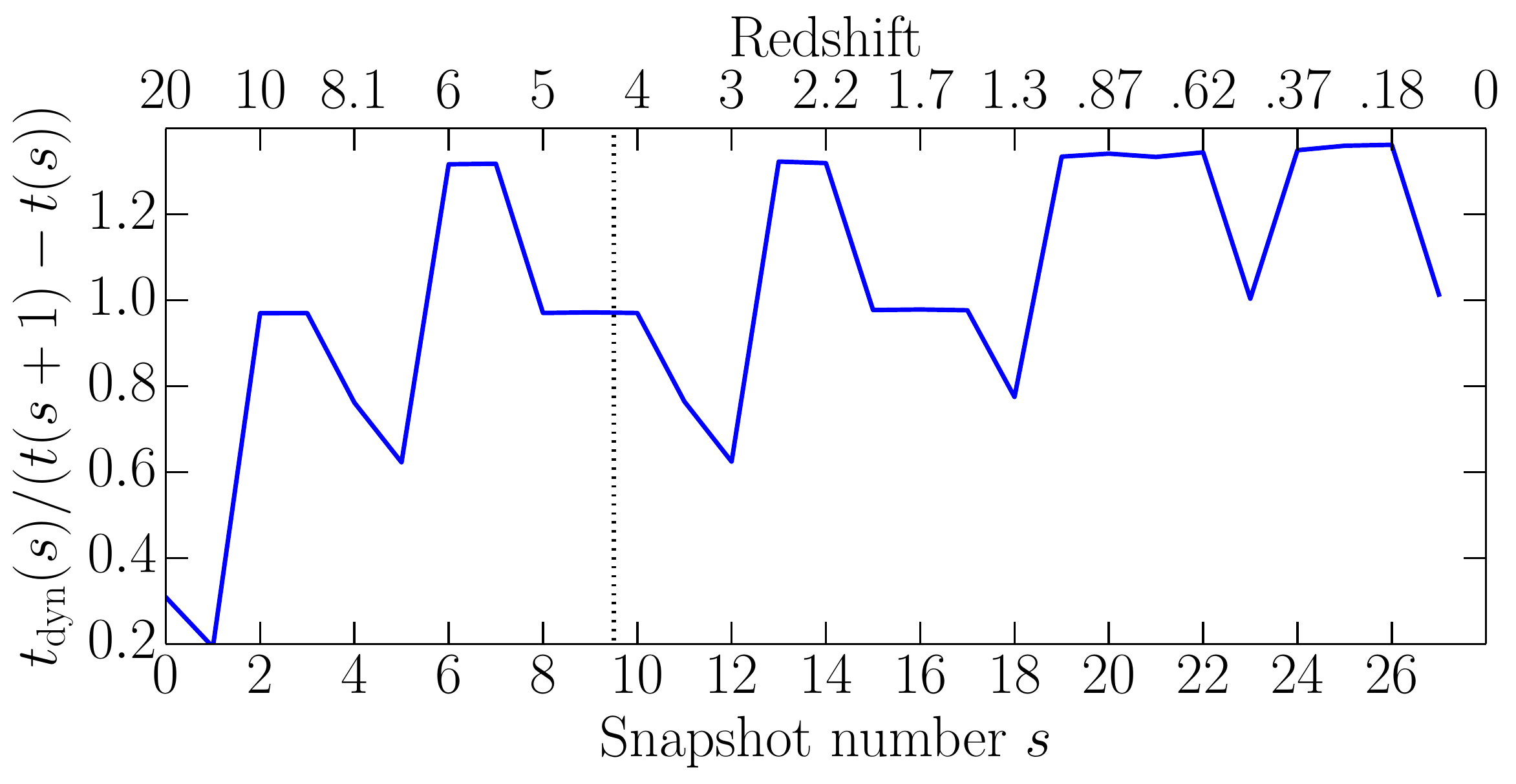}
	\caption{Ratio of a halo dynamical time at the redshift of snapshot number $s$ to the time to the next snapshot in EAGLE.  Snapshots to the right of the vertical dotted line were included in the analysis of this paper.}
	\label{fig:timesep}
\end{figure}

A cooling episode should occur over a time-span comparable to the dynamical time of a halo, which is directly tied to the Hubble flow at the given redshift:
\begin{equation}
t_{\rm dyn} = R_{\rm vir} / V_{\rm vir}= \left[10\, H(z) \right]^{-1}\, ,
\label{eq:tdyn}
\end{equation}
where $R_{\rm vir}$ and $V_{\rm vir}$ are the virial radius and velocity, respectively, of a given halo.  As we want to capture the before and after states of a cooling episode, snapshots separated by a dynamical time would be well suited for our study.  By no coincidence, the temporal separation between snapshots in EAGLE never differs from a dynamical time by more than $\sim$40\% (with the exception of the first two snapshots in the simulation, which we do not use in this paper).  This is illustrated by Fig.~\ref{fig:timesep}.

\begin{figure}
	\centering
	\includegraphics[width=0.95\textwidth]{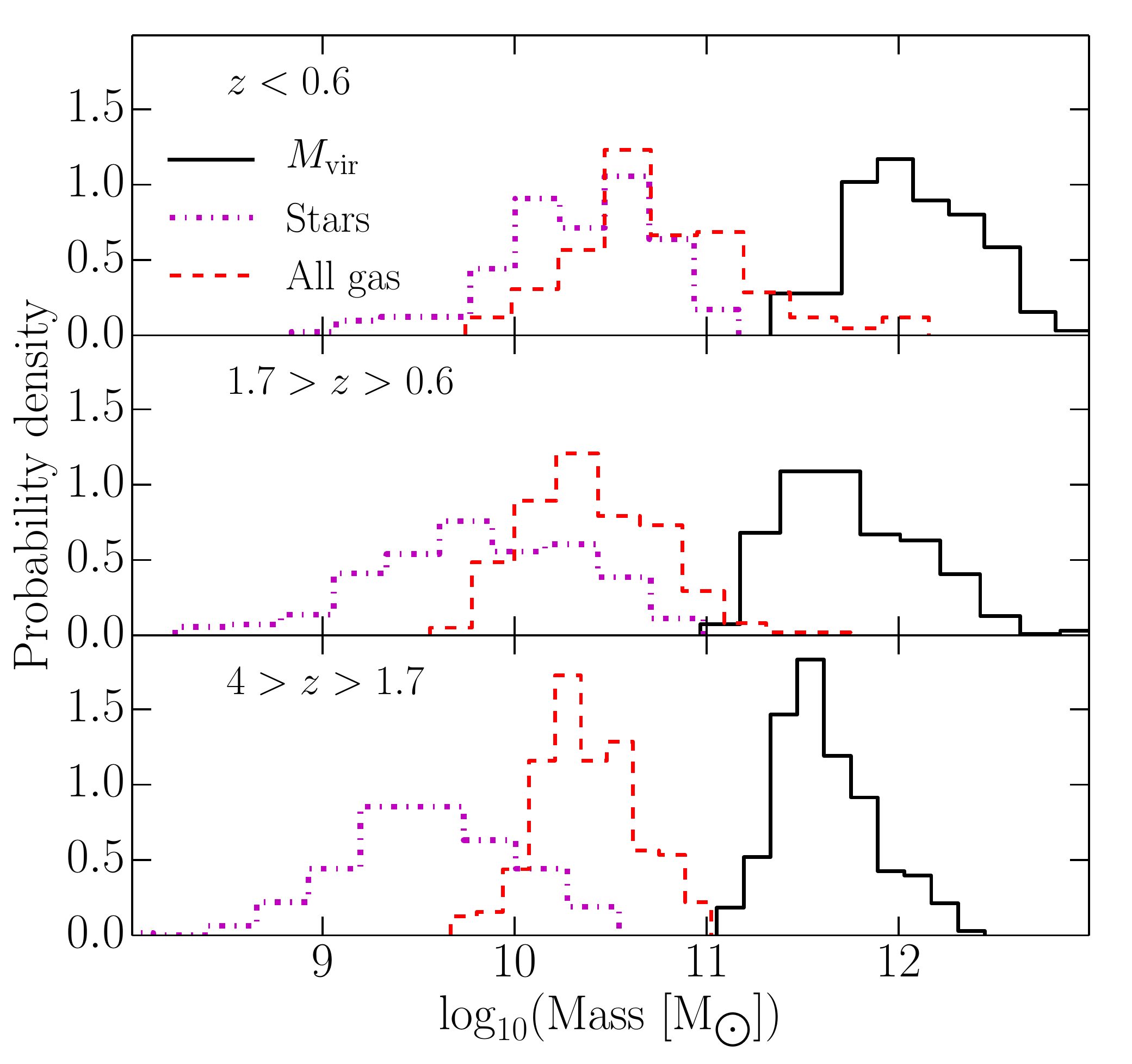}
	\caption{Normalised histograms of masses of Reference EAGLE systems used in this paper, according to the selection criteria given in Section \ref{ssec:sample}.  The solid distributions give the haloes' virial masses.  The dashed and dot-dashed distributions give the gas and stellar masses, respectively, of all particles in the main subhalo within $R_{\rm vir}$.  Each panel covers a different redshift range, as labelled, with the same number of snapshots.}
	\label{fig:sample}
\end{figure}

The precise threshold imposed for the minimum number of cooling particles is, at some level, arbitrary.  A threshold is required to avoid numerical noise and to ensure a meaningful mass cools.  We find a round value of 64 particles returns a sizeable number of haloes (an average of 48 per snapshot interval in the 25-Mpc box), each with sizeable cooling rates ($\gtrsim$ 0.2 ${\rm M}_{\odot}\,{\rm yr}^{-1}$).
Of course, the true cooling rates of haloes over each snapshot interval can be higher, as there may have been particles that cooled and were reheated within the interval that we missed.  It is possible then that our results are biased towards higher-mass systems with high cooling rates, and caution should be heeded when extrapolating to lower mass (see Fig.~\ref{fig:sample} and below).  By comparing the recorded historical maximum temperature of gas particles before and after cooling in our sample, we know that the ratio of hot particles that cool and are reheated between snapshots is, in the majority of cases, small compared to those that cool and remain cold, but it is difficult to quantify exactly which particles have been affected by feedback (beyond those directly injected with energy) with snapshot data.
As will be shown, our 64-particle threshold leads to results that are consistent with the RecalHR simulation when we select the same haloes.  We are, however, restricted to studying haloes at $z<4$.  The average temporal separation between snapshots in this range is 681 Myr.

We present the masses of our sample of Reference EAGLE haloes in Fig.~\ref{fig:sample}.  We break these into three redshift bins, each with 6 snapshot pairs, which we use throughout the paper.  The normalised histograms show the virial masses of the haloes, as well as the gas and stellar masses of all particles in each respective main subhalo inside an aperture of $R_{\rm vir}$.  Note that the gas masses are for \emph{all} gas, i.e.~hot + cold + everything in between.

\section{Before cooling: hot gas in haloes}
\label{sec:hot}
\subsection{Temperature profiles of hot gas}
\label{ssec:temphot}
In the absence of cold streams, before finding itself in a galaxy, gas will find itself in a hot state, sitting in a halo.  Models of galaxy evolution often operate under the assumption that hot gas in haloes, and therefore hot gas that will cool, carries a uniform temperature that is equal to the virial temperature.  To compare against this, we examine the temperature profiles of haloes in our EAGLE sample, each normalised by its virial temperature.  The virial temperature is given by
\begin{equation}
T_{\rm vir} = \frac{\bar{\mu} m_p}{2k_B} V_{\rm vir}^2 = 35.9 \left(\frac{V_{\mathrm{vir}}}{\mathrm{km\,s}^{-1}} \right)^2~\mathrm{K}~,
\label{eq:T_vir}
\end{equation}
where $k_B$ is the Boltzmann constant, and we have adopted a mean molecular weight, $\bar{\mu} m_p$, of 59\% that of the proton mass.

\begin{figure*}
	\centering
	\includegraphics[width=0.95\textwidth]{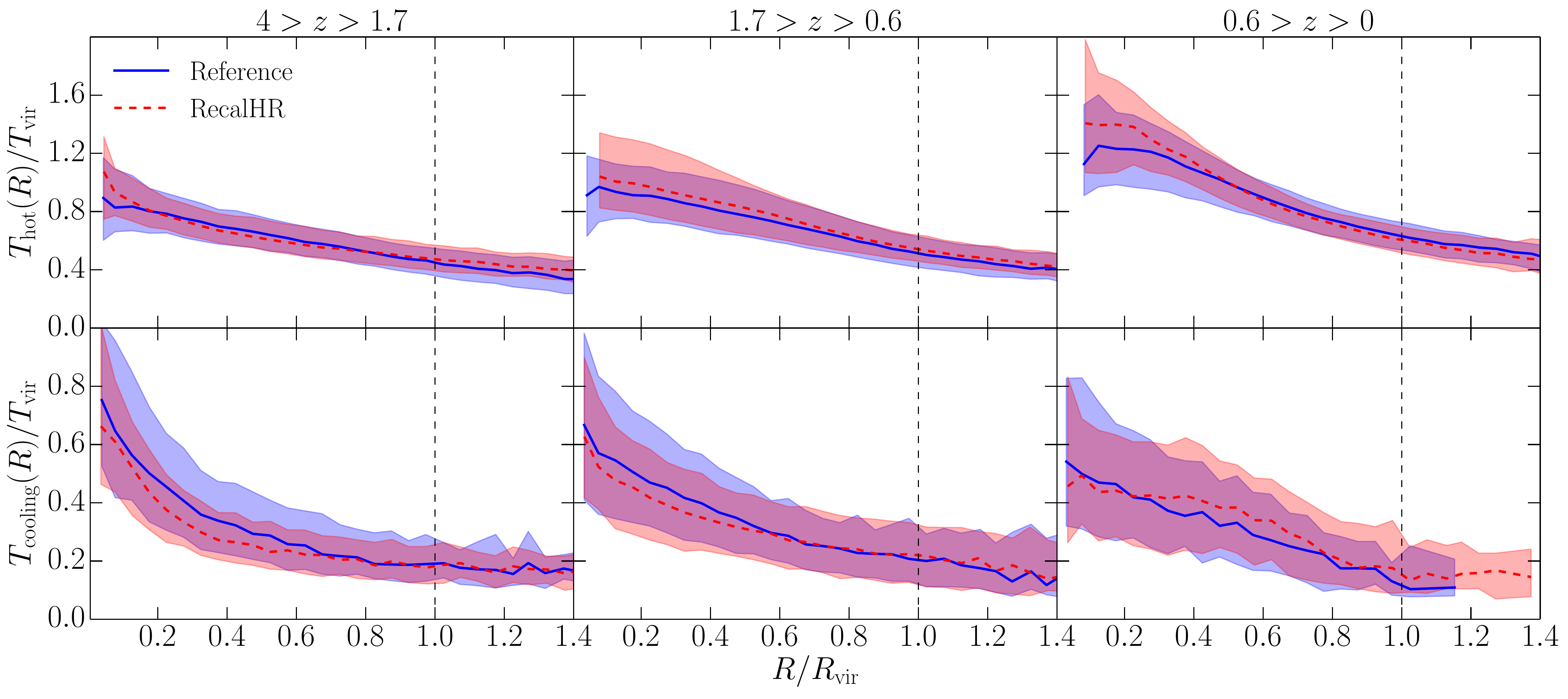}
	\caption{Temperature profiles of non-star-forming, hot ($T>10^5\,{\rm K}$) gas in haloes in the Reference and RecalHR runs of EAGLE.  The top row of panels presents the temperature profiles for \emph{all} hot gas in the haloes, while the bottom panels only include hot gas particles that are known to cool (onto the galaxy) by the following snapshot.  The temperature in both these cases is normalised by the virial temperature of each halo.  Columns of panels are for different redshift ranges.  The median (solid and dashed curves) and inner 68\% (shaded regions of corresponding colour) of profiles are shown in each panel for the two simulations.  Vertical dashed lines are given at $R_{\rm vir}$ to indicate the cut-off for what is typically regarded as part of a halo.  While the hot-gas temperature rises in the centre, the gas that cools continues to get cooler.}
	\label{fig:tempref}
\end{figure*}

We present temperature profiles of hot gas from our sample of EAGLE haloes from the Reference and RecalHR runs in the top panels of Fig.~\ref{fig:tempref}.  The profiles were built by measuring the mean temperature of hot gas particles within spherical shells.  We use (a maximum of) 100 shells for each halo, where each shell encompasses the same number of particles for a given halo, unless that width is below the spatial resolution (softening scale) of the simulation.  We show the evolution of these profiles by grouping systems into three redshift bins, where each bin includes the same number of snapshots.  We display profiles for the median and $16^{\rm th}$--$84^{\rm th}$ percentile range of our samples in Fig.~\ref{fig:tempref}.  It is immediately clear that the hot gas is not truly isothermal, but rather temperature tends to decrease with radius.  At higher redshift, the gas is within a factor of $\sim$2 of being isothermal and only approaches $T_{\rm vir}$ towards the centre.  As time evolves, the centres become hotter and the temperature gradient steepens.

The bottom panels in Fig.~\ref{fig:tempref} consider only the hot gas that will cool (onto the galaxy) before the subsequent snapshot, which we dub `cooling' particles.  As these particles only constitute a small fraction of the hot gas, we reduce the number of spherical shells to 20 in each halo for measuring these profiles.  $T_{\rm cooling}(R)$ has a clear gradient from high redshift onward.  This shows that gas that settles onto a galaxy from a larger distance tends to already be cooler than the inner gas.  Physically, this makes sense, as gas at large radii is at lower density (see below) and hence has less opportunity to lose its energy through collisions over the same time.  It is not surprising that the average temperature of cooling gas is consistently below the hot gas in general; the hotter particles would require longer to cool.  Perhaps what is more interesting is that as the centres of haloes heat overall, the temperature of the \emph{cooling} particles \emph{decreases}.  As discussed further in Section \ref{ssec:rhohot}, the density of hot gas decreases at the centres of haloes over time (Fig.~\ref{fig:rhohot}).  This means there is less opportunity for hot gas to cool through collisions, and thus the temperature difference of a particle before and after a cooling episode (over a time-step) decreases.  Small differences are seen between the Reference and RecalHR runs (which we come back to briefly in Section \ref{ssec:relj}), but they are broadly consistent with each other.

\begin{figure}
	\centering
	\includegraphics[width=0.95\textwidth]{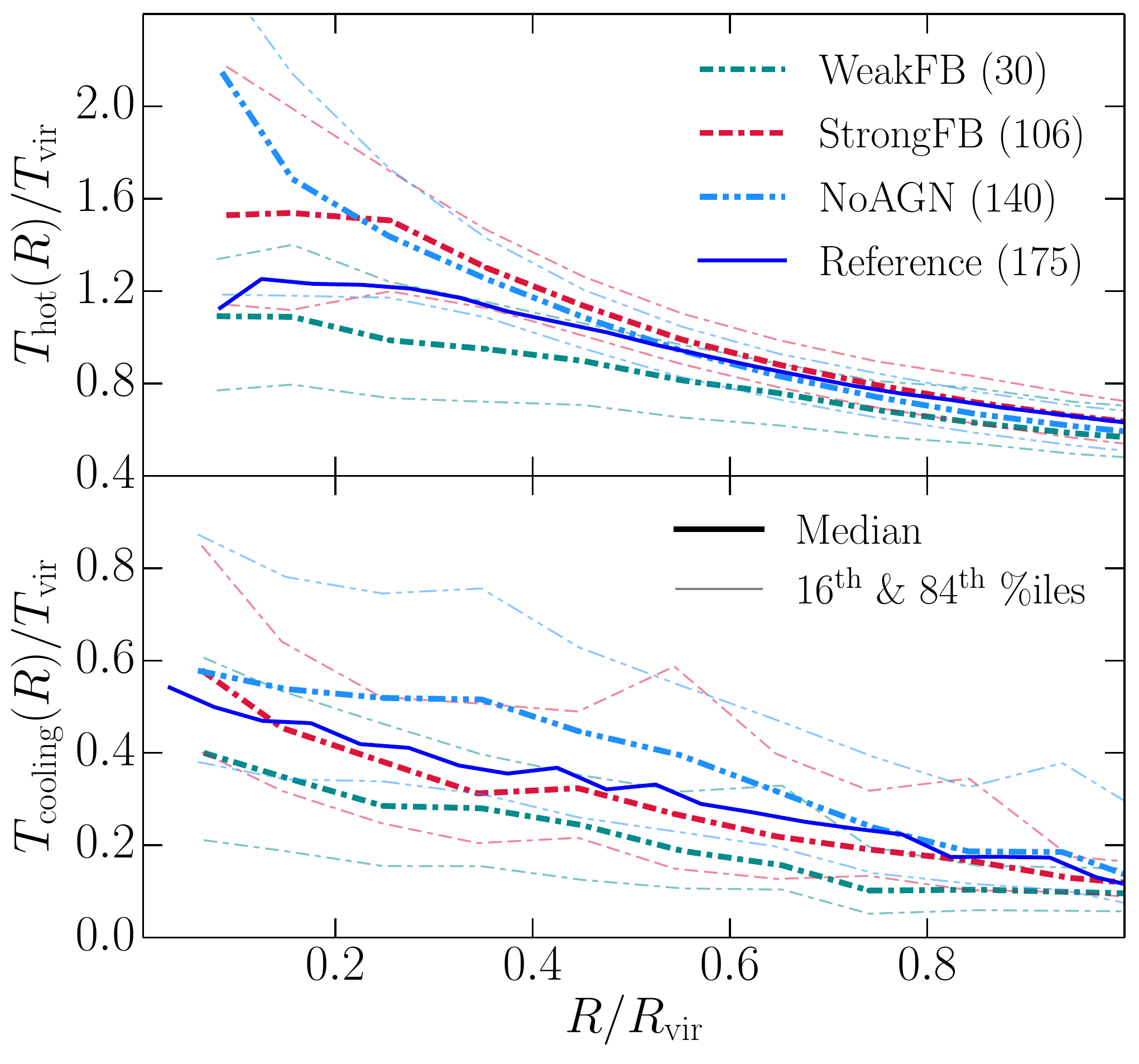}
	\caption{Temperature profiles of hot gas in haloes for various alternate-feedback runs of EAGLE, only considering systems at $z<0.6$.  The top panel includes all hot gas in the profiles, while the bottom panel only includes that which is known to cool before the next time-step.  The thick, opaque curves give the median profiles of the different EAGLE runs.  The thin, transparent lines give the 16$^{\rm th}$ and 84$^{\rm th}$ percentiles for the alternate-feedback runs.  The bracketed numbers in the legend indicate the number of haloes in the sample of each run.  Both stellar and AGN feedback affect the temperature profiles of hot haloes, but the gas in the process of cooling is not affected identically.}
	\label{fig:tempalt}
\end{figure}

To quantify the effect of feedback on the temperature structure of the hot haloes, we have calculated the same temperature profiles for the alternate-feedback runs of EAGLE.  We present these for the lowest redshift bin in Fig.~\ref{fig:tempalt}, comparing the median profile of the Reference simulation to the median, 16$^{\rm th}$ and 84$^{\rm th}$ percentiles of the WeakFB, StrongFB and NoAGN runs.  At higher redshift, the profiles of the alternate-feedback runs more closely matched those of the Reference run (as they have the same initial conditions, and so we have omitted them).  The differences seen at low redshift are hence directly the result of feedback associated with galaxy evolution.  Note that the number of systems meeting our selection criteria in each run varies as a result of the differing feedback.  This is summarised in Table \ref{tab:Nhalo}.  For our results, we only include the haloes in each sample that are also present in the Reference sample.

\begin{table}
\centering
\begin{tabular}{l r r} \hline
Simulation & $N_{\rm halo}$ & $N_{\rm common}$\\ \hline
Reference &  175 & 175 \\
WeakFB & 57  & 30 \\
StrongFB & 106 & 106 \\
NoAGN & 292 & 140 \\ \hline
\end{tabular}
\caption{Number of haloes at $z<0.6$ in each sample from the various runs of EAGLE.  $N_{\rm halo}$ is the full number that met our selection criteria (Section \ref{ssec:sample}).  $N_{\rm common}$ is the number of these haloes that are also in the Reference sample.  Only the latter are used in this paper.}
\label{tab:Nhalo}
\end{table}

Let us first examine the role of stellar feedback.  Stellar feedback is implemented thermally in EAGLE.  This raises the temperature of gas particles in the centre of the haloes, where the exploding stars are.  These hot particles rise and can drive shocks that heat their neighbours.  In our sample, the masses of the haloes are large enough that supernova energy is insufficient to eject gas from galaxies out of the halo entirely.  The result instead is that the gradient of the temperature profile becomes generally steeper.  As seen in the top panel of Fig.~\ref{fig:tempalt}, the WeakFB run shows lower $T_{\rm hot}(R)$ than the Reference simulation across all radii and has a shallower gradient.  Meanwhile, the StrongFB run has a stronger gradient, and is notably hotter towards the centre.  

Comparison of the WeakFB and Reference simulations in the bottom panel of Fig.~\ref{fig:tempalt} highlights that less efficient stellar feedback leads to less-massive and less-turbulent hot-gas haloes at low redshift, which reduces opportunity for cooling, thus lowering $T_{\rm cooling}(R)$.  At higher redshift, cooling is still efficient, and due to a lack of reheating of this cooled gas, the cooling rates of the WeakFB haloes at low redshift are reduced relative to the Reference run, and hence there are far fewer galaxies in the former's sample (see Table \ref{tab:Nhalo}).  Moving from the Reference to the StrongFB run does not change $T_{\rm cooling}(R)$ significantly at low $z$, however.  While there is more hot gas in the case of StrongFB, most of it is at a high enough temperature such that it may not mix effectively and does not have sufficient time to cool over a time-step.  The temperature signature of the cooling gas hence remains unchanged.  Instead, due to the haloes being hotter, we find fewer StrongFB haloes are actively cooling at low redshift versus the Reference run (but more than the WeakFB case), as highlighted in Table \ref{tab:Nhalo}.

Let us now consider the role of AGN.  Perhaps counter-intuitively, switching AGN off leads to an evolution in $T_{\rm hot}(R)$ that looks similar to the StrongFB case (top panel of Fig.~\ref{fig:tempalt}).  While AGN feedback does heat the central gas particles in a halo, the net effect is less a case of the temperature simply rising at the centre, and more a case of AGNs regulating the temperature of the gas throughout the haloes.  Strong outflows from AGNs advect hot particles to larger distances from the galaxy than stellar-driven outflows would, thereby spreading the heat out and reducing the temperature gradient.  In fact, for the haloes common to both the Reference and NoAGN samples, we found that the total baryonic mass within $R_{\rm vir}$ for the NoAGN run was higher by $\sim$0.2--0.3 dex for the low-redshift bin, implying that AGN outflows were able to eject baryons from the haloes.  This also helps to explain why there are 67\% more NoAGN haloes that are actively cooling at $z<0.6$ compared to the Reference run (see Table \ref{tab:Nhalo}).

The difference between the NoAGN run and the others is that $T_{\rm cooling}(R)$ also increased throughout the halo (see the bottom panel of Fig.~\ref{fig:tempalt}).  This shows that an AGN suppresses the ability of hot gas to cool, as now, without AGN, hotter gas is seen to cool more efficiently.  This is precisely how radio mode AGN feedback is considered in semi-analytic models \citep[e.g.][]{bower06,croton06}; that is, the heating rate of a radio AGN is subtracted directly off the cooling rate of a halo to calculate its net cooling rate (see Section \ref{app:rates}).  Although, note that EAGLE does not distinguish between radio and quasar modes of AGN, instead opting for a single model of AGN feedback which is more in line with the quasar mode \citep[see][]{schaye15}.  Even if the net effect of AGN feedback in EAGLE is to make the cooling process less efficient, one would expect the properties of AGN-driven winds in EAGLE to differ if radio mode feedback were implemented directly in the simulation \citep[see][]{dubois12}.

If one is to believe the results of the EAGLE simulations, it seems appropriate for semi-analytic models to treat the effects of stellar and AGN feedback on cooling separately.  To summarise more specifically, in Milky Way-mass haloes in EAGLE, stellar feedback heats cold gas but does not prevent fresh gas from cooling, whereas AGN feedback can eject gas from the halo entirely and more directly impacts cooling gas by regulating the temperature structure of the halo, which can suppress the efficiency of cooling.
Ideally, prescriptions for feedback effects on cooling in semi-analytic models would be based on observations.  Recent efforts to identify signatures of feedback effects on stacked X-ray observations of cooling hot gas around galaxies have shown consistency with gentle, self-regulated mechanical AGN feedback in the luminosity--temperature relation \citep{anderson15}.  Cold-phase studies of the circumgalactic media of isolated galaxies have found evidence for replenishment through outflows from stellar feedback \citep{nielsen16}, although how this might translate into a model for feedback-affected cooling is yet unclear.
As of now, hydrodynamic simulations are the best tool available for developing coarser models of feedback-affected cooling.  While beyond the scope of this paper, we suggest the EAGLE simulations would provide a good testbed for this follow-up study.

\subsection{Density profiles of hot gas}
\label{ssec:rhohot}

\begin{figure*}
	\centering
	\includegraphics[width=0.95\textwidth]{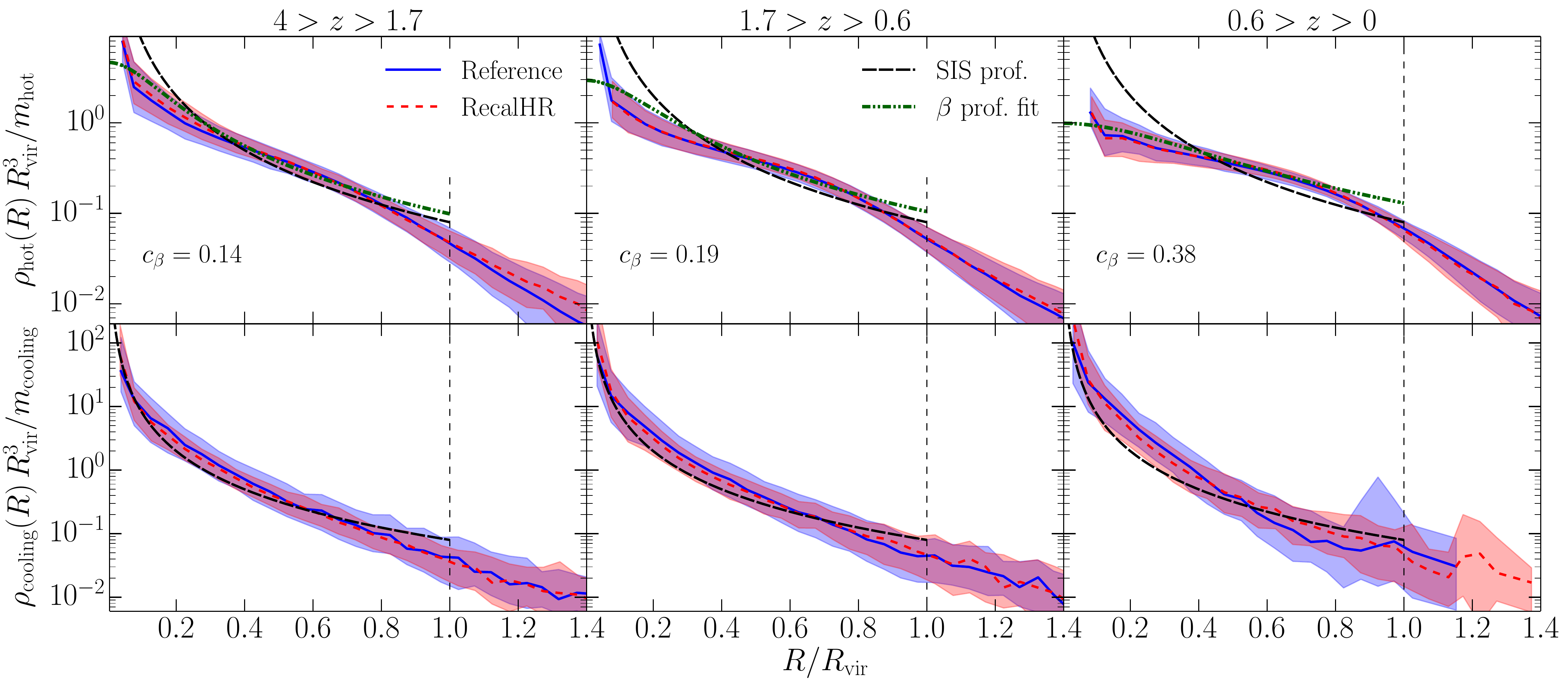}
	\caption{As for Fig.~\ref{fig:tempref} but instead showing density profiles of hot and cooling gas, normalised by the average density within $R_{\rm vir}$.  These are compared against a singular isothermal sphere (SIS) profile in all panels, and against a $\beta$ profile for all hot gas, which is fitted to the median density profile of the Reference simulation in each case.  The numbers in the bottom left of those panels give the value of $c_{\beta}$ for the fit (see equation \ref{eq:rho_beta}).  A fitted $\beta$ profile describes the hot gas as a whole reasonably well, yet the gas that cools is roughly consistent with a singular isothermal sphere.}
	\label{fig:rhohot}
\end{figure*}

We now turn our attention to the density profiles of hot gas in haloes.  Using the same spherical shells as used for the temperature profiles, we measure the density within each shell of our EAGLE haloes.  We normalise the density profile of each hot halo by $m_{\rm hot} / R_{\rm vir}^3$.  We then make the same measurements for the cooling gas, instead normalised by $m_{\rm cooling} / R_{\rm vir}^3$.  This allows us to directly compare the density profiles of both the hot and cooling gas, regardless of the halo size or total hot/cooling mass.  In Fig.~\ref{fig:rhohot}, we present these for the same 3 redshift bins as before.  The top row shows the density profiles for all hot gas, while the bottom row is for cooling gas.  The Reference and RecalHR runs produce consistent profiles with each other, and any differences to the alternate-feedback runs are similarly small on a logarithmic scale (and hence we have not included the profiles for those runs).

A clear evolution is seen in the hot-gas density profiles, where the gradient at the centre becomes shallower with time.  The cooling-gas density profile evolves less noticeably, however, and remains steep.  The fact that cooling is always more effective at the centres of haloes helps elucidate why the total hot-gas density profile forms a core over time.  While the hot gas at the centre becomes hotter and less dense (relative to the outer parts of the same halo), the profile of the gas that actually cools is almost unaffected by feedback emanating from the galaxy.  As mentioned in Section \ref{ssec:temphot}, particles nearer the centre of the galaxy will not need as much time to cool as those at the outskirts, so even if $\rho_{\rm hot}(R)$ were flat, we should still expect a negative gradient for $\rho_{\rm cooling}(R)$.

Some models of galaxy formation assume hot gas follows the density profile of a singular isothermal sphere \citep[e.g.][]{croton06,sage,lagos08,somerville08,lee13,stevens16}, where
\begin{equation}
\rho_{\rm hot}(R) = \frac{m_{\rm hot}}{4 \pi R_{\rm vir} R^2}~.
\label{eq:rho_SIS}
\end{equation}
This relation is over-plotted in each panel in Fig.~\ref{fig:rhohot}.  This profile is based primarily on simplicity, where there is an absence of cited, explicit physical motivation in the models.
An alternative, motivated by both theory and observations of galaxy groups and clusters, is a $\beta$ profile \citep[see the review by][]{mulchaey00}.  This has been incorporated in the \textsc{galform} family of models \citep[e.g.][]{cole00,benson03,font08}.  The $\beta$ profile prescribes the gas density as
\begin{equation}
\rho_{\rm hot}(R) = \rho_0 \left[ 1 + \left(\frac{R}{R_c}\right)^2 \right]^{-3\beta/2}~,
\end{equation}
where $\rho_0$ is the central density and $R_c$ is a core radius.  Under the assumption that all haloes follow the same value of $\beta$ and that $R_c = c_{\beta} R_{\rm vir}$, where $c_{\beta}$ is a constant \citep[e.g., as assumed by][]{benson03}, we can compare our EAGLE profiles to the $\beta$ profile.  Taking $\beta = 2/3$, the expression becomes
\begin{multline}
\rho_{\rm hot}(R) = \frac{m_{\rm hot}}{4 \pi c_{\beta}^2 R_{\rm vir}^3} \left[1 - c_{\beta} \tan^{-1}\left(\frac{1}{c_{\beta}}\right) \right]^{-1} \\ \left[1+\left(\frac{R}{c_{\beta} R_{\rm vir}}\right)^2 \right]^{-1}~.
\label{eq:rho_beta}
\end{multline}
Using the median density profiles of the Reference simulation in Fig.~\ref{fig:rhohot}, we have fitted for $c_{\beta}$ and over-plotted for comparison.

Fig.~\ref{fig:rhohot} clearly shows that singular isothermal spheres are not representative of the distribution of hot gas in our haloes. While the $\rho_{\rm cooling}(R)$ profiles are closer to being exponential on average than anything else, curiously, they do encompass the singular isothermal sphere profile within their 68\% confidence range, at least for $z>0.6$.  A singular isothermal sphere describes the cooling-gas density profile as well as a best-fitting $\beta$ profile would (i.e.~the best-fitting $c_{\beta}$ is large).  Despite not actually having isothermal haloes, the manner in which gas cools onto galaxies in EAGLE is consistent with randomly drawing particles out of a singular isothermal sphere, which we have found is true for the alternate-feedback EAGLE runs as well (not shown here).  To rephrase this, if one's model for halo gas cooling \emph{only} requires a description of the density profile for which hot gas cools out of, then a singular isothermal sphere appears to be a valid approximation for that profile.  In reality, a complete consideration of halo gas cooling in a semi-analytic model tends to require modelling the full hot-gas density distribution, and not just the cooling gas (see Section \ref{ssec:tcool}).

\begin{figure}
	\centering
	\includegraphics[width=0.95\textwidth]{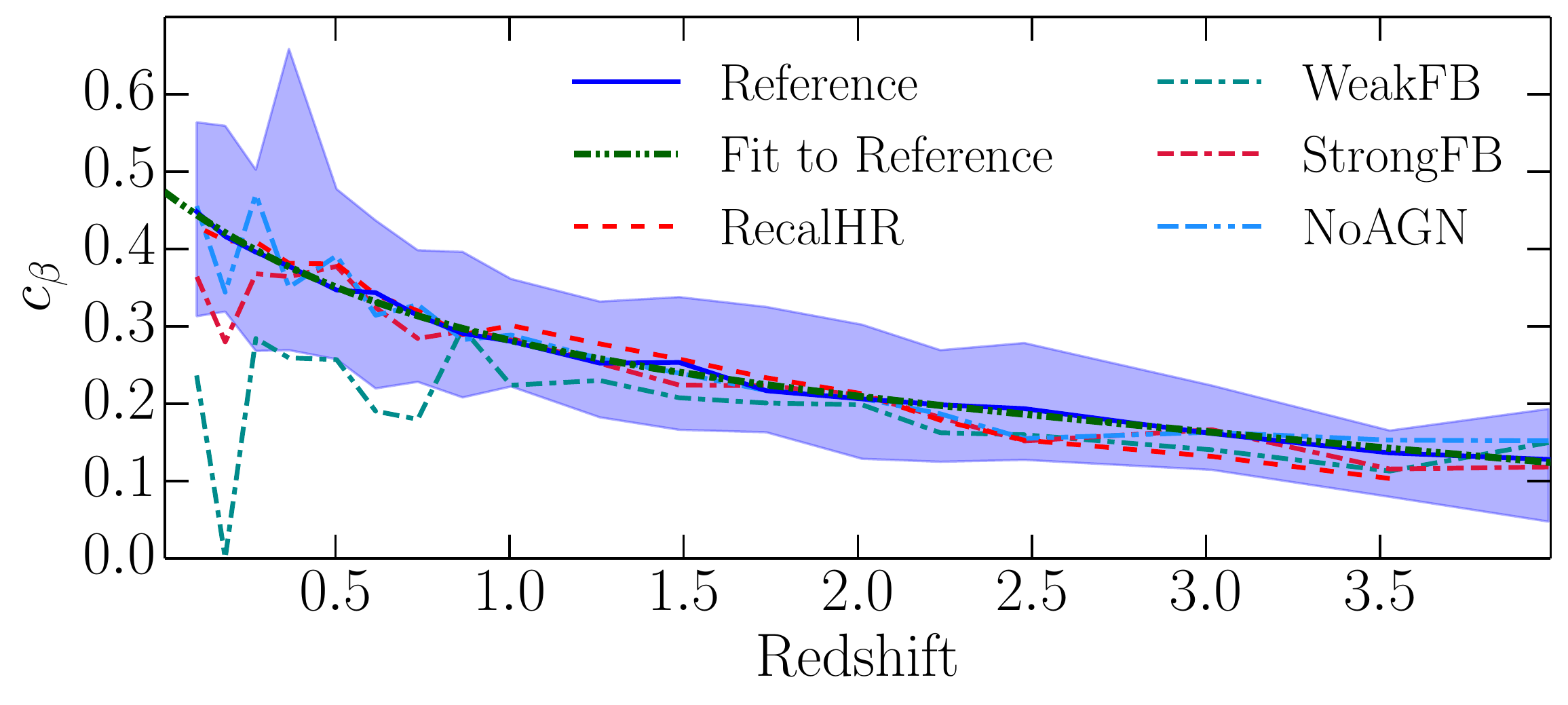}
	\caption{Best-fitting concentration parameter of $\beta$ profile fits to the hot-gas density profiles of haloes in our EAGLE sample at each snapshot, assuming $\beta = 2/3$, for various runs as given in the legend.  Each curve gives the median relation from each simulation, with the exception of the perfectly smooth curve, which is a fit to the Reference simulation median (equation \ref{eq:cfit}), which happens to describe many of the other runs well too.  The shaded region covers the 16$^{\rm th}$--84$^{\rm th}$ percentile range for the Reference simulation.}
	\label{fig:cbetaz}
\end{figure}

A $\beta$ profile captures the total hot-gas density reasonably well out to $\sim$0.8$R_{\rm vir}$, when $c_{\beta}$ is free, as seen in the top panels of Fig.~\ref{fig:rhohot}.  By allowing this, we find the best-fitting $c_{\beta}$ decreases with time.  To quantify this, we have fitted $c_{\beta}$ to each hot-gas density profile of our sample haloes for each snapshot individually.  We have done this for the Reference, RecalHR, and alternate-feedback EAGLE runs, and present the evolution of $c_{\beta}$ for these in Fig.~\ref{fig:cbetaz}.  We include the median relation for all runs, and include the scatter for the Reference simulation.  We find that weak stellar feedback can affect the value of $c_\beta$ at $z \lesssim 1$, but otherwise the fit is fairly robust to feedback changes.  In addition, we find the median $c_{\beta}(z)$ curve for the Reference simulation is fitted almost perfectly by a simple analytic expression:
\begin{equation}
c_{\beta}(z) \simeq 0.20 e^{-1.5 z} - 0.039 z + 0.28~.
\label{eq:cfit}
\end{equation}
This least-squares fit (which weights each snapshot equally) is included in Fig.~\ref{fig:cbetaz}.  We note that for all $z<4$, the best-fitting $c_\beta$ is above the values assumed in previous incarnations of \textsc{galform}: 0.07 \citep{benson03} and 0.1 \citep{font08}.

If we consider all snapshots at once, then we naturally find a weak correlation between $c_{\beta}$ and $M_{\rm vir}$ for our EAGLE haloes.  However, this is purely a result of our mean sample $M_{\rm vir}$ increasing at lower redshift (Fig.~\ref{fig:sample}).  At fixed redshift, there is no clear trend between $c_{\beta}$ and either $M_{\rm vir}$ or $M_{\rm hot}$.  As best evidenced by Fig.~\ref{fig:rhohot}, the cores in these haloes, i.e.~higher $c_{\beta}$ values, come from cumulative cooling, dominant at halo centres.  For our EAGLE haloes, we find $c_{\beta}$ to be strictly a function of redshift and cannot meaningfully find a secondary dependence on any integrated halo property (that a semi-analytic model would have access to).

\subsection{Metallicity profiles of hot gas}
\begin{figure}
	\centering
	\includegraphics[width=0.95\textwidth]{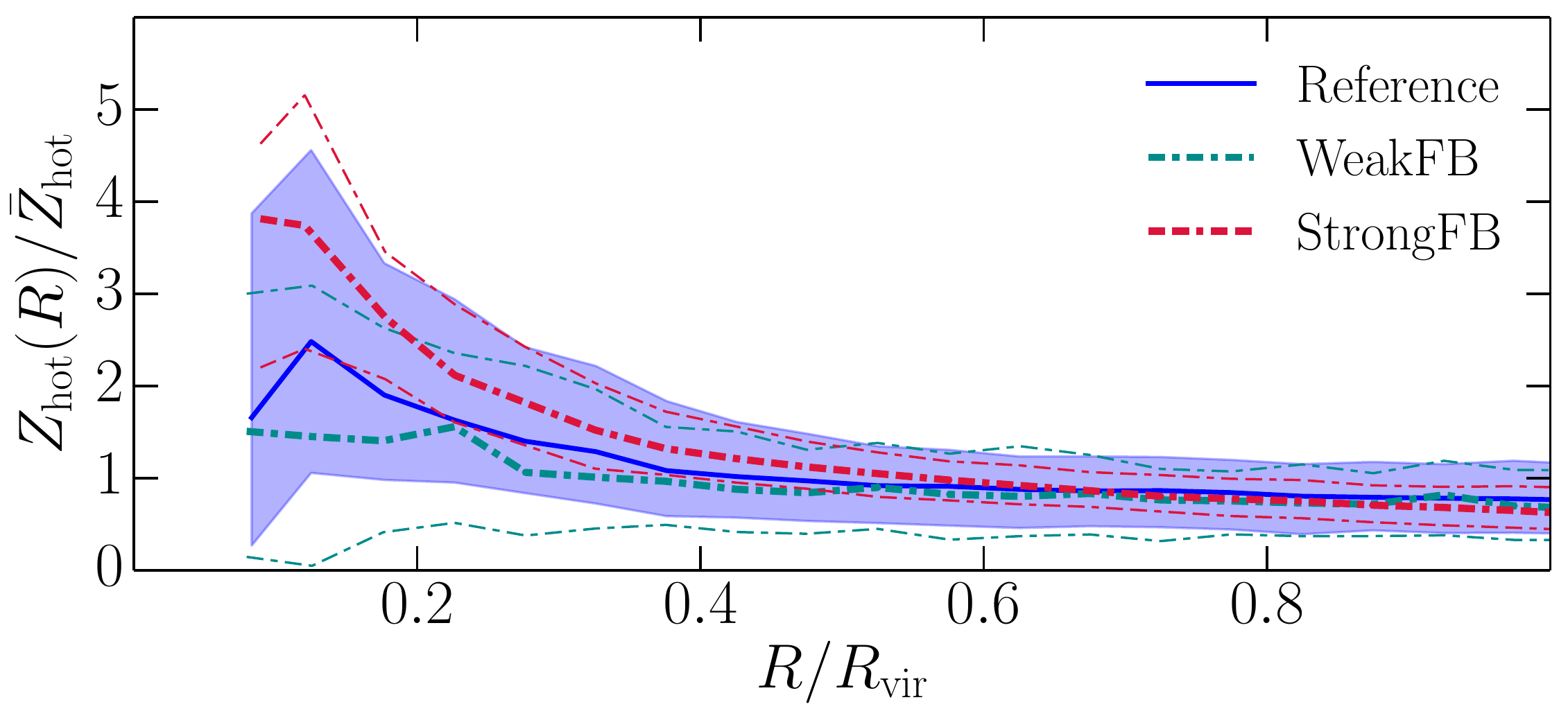}
	\caption{Metal fraction profiles for hot gas in our EAGLE haloes at $z<0.6$, normalised by the average metal fraction in each case.  The thick curves give the median for each labelled simulation.  The shaded region covers the 16$^{\rm th}$--84$^{\rm th}$ percentile range for the Reference simulation, while the thin curves cover the same range for the WeakFB and StrongFB runs.  Stellar feedback is seen to enrich the centres of hot haloes with metals.}
	\label{fig:met}
\end{figure}

Another common approximation made in analytic models is that the metallicity of hot gas is uniform throughout the halo.  In reality, we would expect hot gas at the centre of haloes to be more metal-rich than at the outskirts, as the stars at the bottom of the potential well are what produce new metals.  To demonstrate the degree to which this is true for EAGLE, we calculate the mass-weighted mean metal fraction of particles in each spherical shell (as defined in Section \ref{ssec:temphot}) of each halo, and plot the normalised metal fraction profiles of our sample haloes at $z<0.6$ in Fig.~\ref{fig:met}.  We include the median and 16$^{\rm th}$--84$^{\rm th}$ percentile range for the Reference, WeakFB, and StrongFB simulations.  At higher redshift, the behaviour of these three simulations is similar to the Reference case presented at low redshift, and thus these have been omitted for simplicity.  The RecalHR simulation is also consistent with the Reference case presented here, and is omitted for clarity.  In all cases, we find metallicity becomes increasingly variant towards the centres of haloes, with a general trend of metallicity increasing.  The effect of stellar feedback enriching the hot halo is clear; by $z<0.6$, the centres of haloes in the StrongFB simulation have the steepest gradients, and the WeakFB simulation the shallowest.  As the rate at which gas cools is dependent on metallicity, metallicity gradients can have an impact on the cooling calculation in a semi-analytic model, which we turn our attention to next.

\subsection{Cooling time profiles}
\label{ssec:tcool}
The primary purpose of semi-analytic models including prescriptions for the hot-gas density profile, temperature, and metallicity is to determine a cooling rate at each time-step.  It is, therefore, of interest what consequences will arise for the cooling rates from the differences in the profiles of the EAGLE haloes versus analytic approximations.  To investigate this, we must first address the typical process for calculating cooling rates in semi-analytic models.  

The majority of semi-analytic models \citep[e.g.][]{cole00,hatton03,cora06,croton06,sage,somerville08,guo11,lee13} calculate cooling rates using some variation of the method presented in \citet{white91}, which is as follows (but see, e.g., \citealt*{monaco07} for a more detailed treatment).  Given a spherically symmetric distribution of hot gas, the `cooling time' for hot gas at a given radius is first defined as
\begin{equation}
t_{\rm cool}(R) \equiv \frac{3}{2}     \frac{T_{\rm hot}(R)}{\rho_{\rm hot}(R)} \frac{\bar{\mu}m_p\,  k_B}{\, \Lambda(T_{\rm hot}(R),Z_{\rm hot}(R))}~,
\label{eq:tcool}
\end{equation}
where $\Lambda(T,Z)$ is the cooling function, commonly drawn from the tables of \citet{sutherland93}, dependent on the temperature and metallicity of the gas.  One then defines the `cooling radius', $R_{\rm cool}$, as the radius at which the cooling time equals a relevant time-scale.  For the purposes of this paper, we have chosen to equate this time-scale to the dynamical time, in line with the {\sc sage} family of models \citep{sage,stevens16,tonini16}, but this could have been informed by the free-fall time-scale instead, as in {\sc galform}, for example.  The general argument then is that the cooling mass that crosses $R_{\rm cool}$ is approximately equal to the cold-gas mass deposition rate onto the galaxy \citep{bertschinger89}.  From this continuity law, one can calculate the rate at which gas cools onto the galaxy as
\begin{equation}
\dot{m}_{\rm cool,model} = 4 \pi\, \rho_{\rm hot}(R_{\rm cool})\, R_{\rm cool}^2 \left( \frac{{\rm d} t_{\rm cool}}{{\rm d} R} \Big |_{R \rightarrow R_{\rm cool}} \right)^{-1}~.
\label{eq:mdot}
\end{equation}
In the case where $R_{\rm cool} > R_{\rm vir}$, haloes are assumed to be in a cold-accretion regime in a semi-analytic model, where the cooling rate is then taken as the ratio of the hot mass to the dynamical (or free-fall) time.  We remind the reader that our EAGLE haloes are selected to be in the hot mode of accretion.

Our intent here is \emph{not} to compare the true cooling rates of EAGLE haloes to those inferred by equation (\ref{eq:mdot}).  To make that comparison would require a complete deconstruction of how feedback influences cooling in both EAGLE and in semi-analytic models.  This is non-trivial, as the way in which feedback and cooling are handled in semi-analytic models is not directly comparable to what goes on in a hydrodynamic simulation, and there is plenty of variation amongst models as well \citep[see][]{lu11}.  A model can have degeneracies when it comes to cooling and heating rates, where the free parameters governing these are only really constrained in a relative sense, such that the \emph{net} growth of (sometimes just the stellar content of) galaxies is representative of the observed Universe.  This is why many semi-analytic--hydrodynamic comparison projects have excluded feedback (and sometimes even star formation) from the simulations entirely \citep[e.g.][]{benson01,yoshida02,helly03,cattaneo07,viola08,saro10,lu11,monaco14}.  What we aim to address here is how the cooling time profiles, through equation (\ref{eq:tcool}), vary when we include the detail of the density, temperature, and metallicity profiles we have available to us from EAGLE.  In Section \ref{app:rates}, we go one step further, and calculate how the $t_{\rm cool}(R)$ profiles of our EAGLE haloes would translate into an effective semi-analytic cooling rate.

\begin{figure*}
	\centering
	\includegraphics[width=0.95\textwidth]{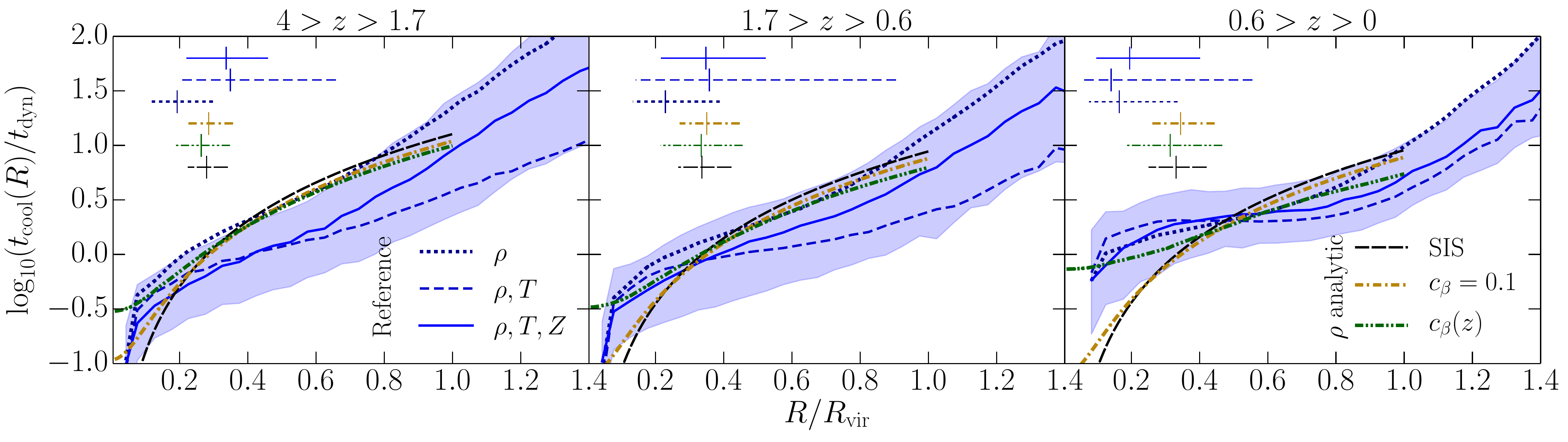}
	\caption{Cooling time of hot gas in our EAGLE halo sample, normalised by their respective dynamical times, as a function of radius, normalised by their virial radii.  We present median profiles calculated from the EAGLE haloes through equation (\ref{eq:tcool}) in cases where we consider (i) only the density profiles of the halo with $T_{\rm hot}(R)=T_{\rm vir}$ and $Z_{\rm hot}(R)=\bar{Z}_{\rm hot}$ (dotted curves), (ii) both the density and temperature profiles with $Z_{\rm hot}(R)=\bar{Z}_{\rm hot}$ (short-dashed curves), and (iii) the density, temperature, and metallicity profiles of the haloes (solid curves).  In the latter case, we show the 16$^{\rm th}$--84$^{\rm th}$ percentile range with the shaded region.  Included for comparison are the median cooling time profiles when assuming the density profile follows (i) a singular isothermal sphere, (ii) a $\beta$ profile with constant $c_{\beta}=0.1$, and (iii) a $\beta$ profile with $c_{\beta}$ calculated from equation (\ref{eq:cfit}), which take $T_{\rm hot}(R)=T_{\rm vir}$ and $Z_{\rm hot}(R)=\bar{Z}_{\rm hot}$ in all 3 cases (see the legend in the right-hand panel).  The horizontal lines cover the inner 68\% of $R_{\rm cool}/R_{\rm vir}$ values in each case, with matching linestyles.  The vertical marks through these give the median cooling radii.  Complete information on the density, temperature, and/or metallicity profiles of hot haloes leads to a greater range of cooling profiles than analytic approximations would give, which can impact cooling radii, thereby affecting the cooling rates one would calculate in a semi-analytic model.}
	\label{fig:tcool}
\end{figure*}

It is normal for the purposes of a semi-analytic model to simplify equation (\ref{eq:tcool}) by taking $T_{\rm vir}$ as the temperature for all hot gas in the halo, and assuming all hot gas to have the same metallicity.  This then leaves $t_{\rm cool}(R)$ for a given halo dependent on the assumed density profile.  We address the impact of the density profile on $t_{\rm cool}(R)$ in Fig.~\ref{fig:tcool}.  The long-dashed and dot-dashed curves give the median $t_{\rm cool}(R)$ profiles after using analytic profiles of a singular isothermal sphere and $\beta$ profiles with $c_{\beta} = 0.1$ and $c_{\beta}(z)$ from equation (\ref{eq:cfit}), respectively.  These profiles differ little, where the $\beta$ fit only makes a notable difference for $R \lesssim 0.3 R_{\rm vir}$ (especially at low redshift), where it is generally true that $t_{\rm cool}<t_{\rm dyn}$.  As a result, the cooling radii calculated from these density profiles are all consistent.  This is shown by the horizontal lines in Fig.~\ref{fig:tcool}, which cover the 16$^{\rm th}$--84$^{\rm th}$ percentile range for $R_{\rm cool}/R_{\rm vir}$ in each case, with the intersecting vertical marks giving the medians.  

As a direct comparison to the analytic profiles, we calculate $t_{\rm cool}(R)$ using the actual density profiles from each Reference EAGLE halo, while maintaining the use of $T_{\rm hot}(R)=T_{\rm vir}$ and $Z_{\rm hot}(R)=\bar{Z}_{\rm hot}$.  The median profile (dotted curves in Fig.~\ref{fig:tcool}) is consistent with the analytic cases in the outer parts of the halo, but diverges for $R \lesssim 0.4 R_{\rm vir}$ for all $z<4$ (the scatter in all these cases is consistent, but this is not shown for clarity).  As a result, the cooling radii are systematically lower than their analytic counterparts.  This highlights the fact that, even when using a density profile that fits the general population by construction, it is difficult to recover the true cooling radii of haloes from a hydrodynamic simulation.  

Next, we consider the role of the temperature profiles of the EAGLE haloes on $t_{\rm cool}(R)$.  We again solve equation (\ref{eq:tcool}) with $Z_{\rm hot}(R)=\bar{Z}_{\rm hot}$, but use the actual profiles of the haloes for $T_{\rm hot}(R)$ and $\rho_{\rm hot}(R)$.  The median $t_{\rm cool}(R)$ profile in this case is given by the short-dashed curves in Fig.~\ref{fig:tcool}.  Comparing this to the dotted curves, we see the profile flattens, which leads to a much broader range in cooling radii for the haloes.  Even though a typical EAGLE halo will only have a radial variation in its temperature around a factor of 3 (cf.~Fig.~\ref{fig:tempref}), this temperature structure can significantly impact the $t_{\rm cool}(R)$ profile one infers for a halo, and therefore would impact the cooling rate one would determine in a semi-analytic model.

As a final step, we include the metallicity profiles of the EAGLE haloes and recalculate $t_{\rm cool}(R)$.  We include the median relation with the solid curves, and the 16$^{\rm th}$--84$^{\rm th}$ percentile range with the shaded region in each panel of Fig.~\ref{fig:tcool}.  The addition of the $Z_{\rm hot}(R)$ profile restores sensible cooling radii values that are consistent with the haloes being in the hot mode of accretion.  For completeness then, a model of halo gas that includes temperature structure should also include metallicity structure for the sake of calculating cooling radii and rates.  For $z>0.6$, these cooling radii are in moderate agreement with the purely analytic profiles.  The $t_{\rm cool}(R)$ profiles flatten at low redshift, which leads to a broad range in cooling radii, which become systematically less than the cases of the analytic profiles.

\subsection{Model-equivalent cooling rates}
\label{app:rates}
\begin{figure*}
	\centering
	\includegraphics[width=0.95\textwidth]{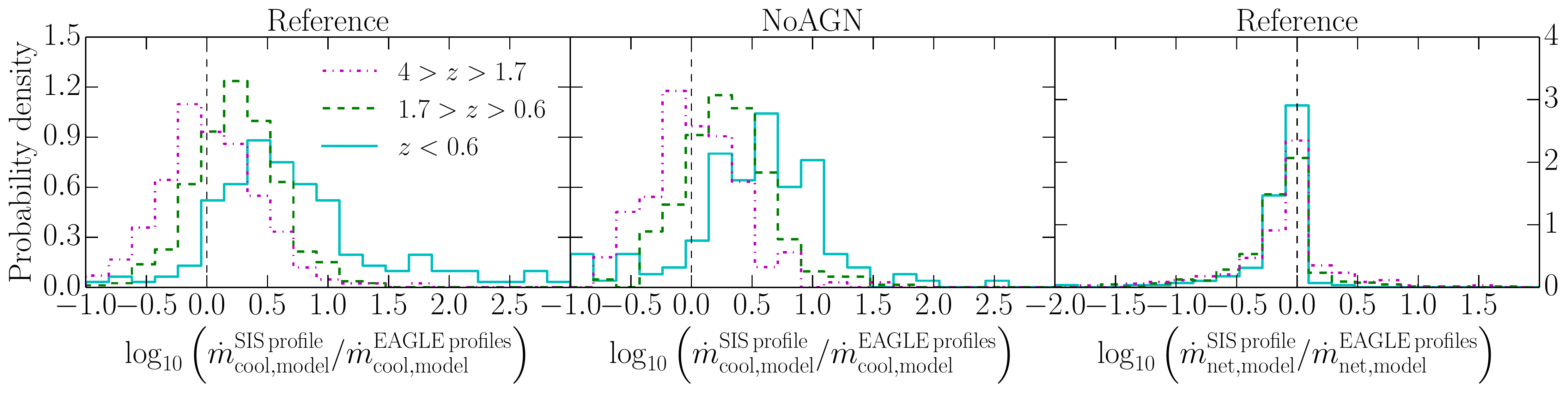}
	\caption{Normalised histograms for the relative analytic cooling rates calculated for model density profiles of hot gas versus those using the actual density, temperature, and metallicity profiles from our sample of Reference and NoAGN EAGLE haloes, as labelled.  The left and middle panels compare gross cooling rates (equation \ref{eq:mdot}), while the right-hand panel compares net cooling rates (equation \ref{eq:mnet}).  If the analytic approximations were sufficient for calculating cooling rates, the distributions would be $\delta$ functions at the vertical dashed line.}
	\label{fig:mcool}
\end{figure*}

In an ideal world, one would know the unique density, temperature, and metallicity profile of hot gas in every halo processed through a semi-analytic model.  By using the measured profiles from EAGLE, we can solve equation (\ref{eq:mdot}) numerically for each halo, thereby determining the `semi-analytic equivalent' for the cooling rate with ideal information.  This can be compared against a more realistic calculation for a semi-analytic model, where in addition to an analytic density profile, it is assumed that $T_{\rm hot}(R)=T_{\rm vir}$ and $Z_{\rm hot}(R)=\bar{Z}_{\rm hot}$ for each halo.  The results of Section \ref{ssec:tcool} suggest any of a singular isothermal sphere, a $\beta$ profile with $c_{\beta}=0.1$, or a $\beta$ profile with $c_{\beta}(z)$ from equation (\ref{eq:cfit}) will work pracitcally the same for calculating a cooling rate.  In the left-hand panel of Fig.~\ref{fig:mcool} we compare the cooling rates calculated by equation (\ref{eq:mdot}) using the full EAGLE profile information against the case of an assumed singular isothermal sphere (we have checked that this is essentially the same for the case of a $\beta$ profile).

The spread in the true $R_{\rm cool}$ values for the EAGLE haloes is greater than that of the analytic models, as seen in all panels of Fig.~\ref{fig:tcool}. This then gives rise to the notable spread in the relative cooling rates seen in the left-hand (and middle, see below) panel of Fig.~\ref{fig:mcool}.  In the lower-redshift bins, the model $R_{\rm cool}$ values are systematically too large, most clearly seen in the right-hand panel of Fig.~\ref{fig:tcool}.  This then translates into systematically higher cooling rates, as seen by the evolution in the distributions of the left-hand (and middle) panel of Fig.~\ref{fig:mcool}.

Note that the cooling rates calculated by equation (\ref{eq:mdot}) do not consider feedback (of any kind), and hence are \emph{gross} cooling rates (as opposed to \emph{net} cooling rates -- see below).  If feedback were to affect the density or temperature profiles of the hot gas in haloes significantly, then we might expect this feedback to still cause differences in the gross cooling rates.  While we showed that feedback does alter the temperature at the centre of EAGLE haloes in Fig.~\ref{fig:tempalt}, this is less than a factor of 2.  Because the hot-gas density profiles also do not change significantly when the strength of feedback is varied, the gross cooling rates calculated by equation (\ref{eq:mdot}) should agree for the Reference and alternate-feedback simulations.  To demonstrate one example, we compare $\dot{m}_{\rm cool}$ for the NoAGN simulation, as we did for the Reference simulation, in the middle panel of Fig.~\ref{fig:mcool}.  While there are small differences compared with the left-hand panel, the evolution of the distributions and their peaks are in agreement. Results for WeakFB and StrongFB are indeed similar, so we omit them for simplicity.

While stellar feedback is often considered independently from calculating cooling rates in semi-analytic models \citep[but not always -- see, e.g.,][]{monaco07}, radio mode AGN feedback is generally implemented by \emph{directly} suppressing the cooling rates rates calculated through equation (\ref{eq:mdot}) \citep[see][]{bower06,croton06}.  Galaxies hosting a larger supermassive black hole will typically have their cooling suppressed more.  At later times, the haloes in our sample are bigger (see Fig.~\ref{fig:sample}) and host heavier black holes.  Because the suppressive heating from AGN need not be tied directly to the gross cooling rates of haloes, ratios of \emph{net} cooling rates as calculated in semi-analytic models with different hot-gas radial profiles will not be the same as the ratio of \emph{gross} cooling rates.  They will, in fact, be systematically smaller for systems with an AGN, effectively counteracting the redshift evolution of the gross cooling rate ratios presented in the left and middle panels of Fig~\ref{fig:mcool}.

To demonstrate the impact AGN heating might have on cooling rates in semi-analytic model, we use an example model of radio mode heating to calculate effective net cooling rates for the haloes.  We calculate the heating rate based on energy released from Bondi--Hoyle accretion \citep{bondi52} of hot gas in the halo
\begin{equation}
\dot{m}_{\rm heat,model} = \frac{15 \pi}{16} \frac{(\bar{\mu} m_p)^2}{\Lambda(T_{\rm vir},\bar{Z}_{\rm hot})} G c^2\, \kappa \eta\,  m_{\rm BH}
\label{eq:mheat}
\end{equation}
\citep[as implemented similarly in][]{croton06,somerville08}, where $m_{\rm BH}$ is the black-hole mass,\footnote{94\% of the (sub)haloes in our sample have more than one black-hole particle.  In a semi-analytic model, galaxies typically are only allowed one black hole; when a merger brings in a new black hole, it is assumed to merge immediately with any pre-existing black hole.  To mimic this, we sum the masses of all black holes in an EAGLE galaxy to give $m_{\rm BH}$ to calculate $\dot{m}_{\rm heat,model}$.} $\kappa$ is a model parameter used to control the strength of feedback, and $\eta$ is the efficiency with which the inertial mass of the gas is released during accretion onto the black hole.  Here, we assume typical values of $\kappa = \eta = 0.1$.  The net cooling rate of gas in a halo can then be found as the difference between equations (\ref{eq:mdot}) and (\ref{eq:mheat}):
\begin{equation}
\dot{m}_{\rm net,model} = \dot{m}_{\rm cool,model} - \dot{m}_{\rm heat,model}~.
\label{eq:mnet}
\end{equation}
We calculate $\dot{m}_{\rm net}$ for our haloes in the Reference simulation, using the true and analytic (singular isothermal sphere and $\beta$) radial profiles.  In the right-hand panel of Fig.~\ref{fig:mcool} we present the ratio of the model-equivalent net cooling rates using the EAGLE radial profiles (density, temperature, and metallicity) to the singular isothermal sphere profile.  Not only is the evolution of the distributions minimised compared to the ratio of gross cooling rates, but the widths of the distributions are also greatly reduced.  The latter would not be true if the central temperature and metallicity of the hot haloes were used in the Bondi--Hoyle model, however. 

In conclusion, while the structural properties of hot gas in haloes can significantly affect gross cooling rates in semi-analytic models, the application of radio mode feedback can compensate for most of this.

\section{Angular momentum of cooling gas}
\label{sec:angmom}
\subsection{Conservation of angular momentum during cooling}
\label{ssec:cons}

An important aspect of most models of gas cooling is the assumption that the angular momentum of the gas is conserved.  Of course, cooling gas can exchange angular momentum with many other parts of the halo in principle, especially through collisions with gas particles not involved in cooling (including those already cold).  Early attempts at forming spiral galaxies with cosmological, hydrodynamic simulations were plagued by an `angular-momentum catastrophe', where gas cooled too quickly and lost too much angular momentum \citep[see, e.g.,][]{katz91,navarro91,navarro97,navarro00}.  Solutions to this problem were found in better resolution \citep{governato04} and more efficient subgrid feedback \citep[e.g.][]{brook11,brook12}.  With these improvements, we are now in a position to use a simulation like EAGLE to make predictions about angular-momentum conservation of cooling gas, or lack thereof, which is informative for analytic models of gas cooling.

\begin{figure*}
	\centering
	\includegraphics[width=0.754\textwidth]{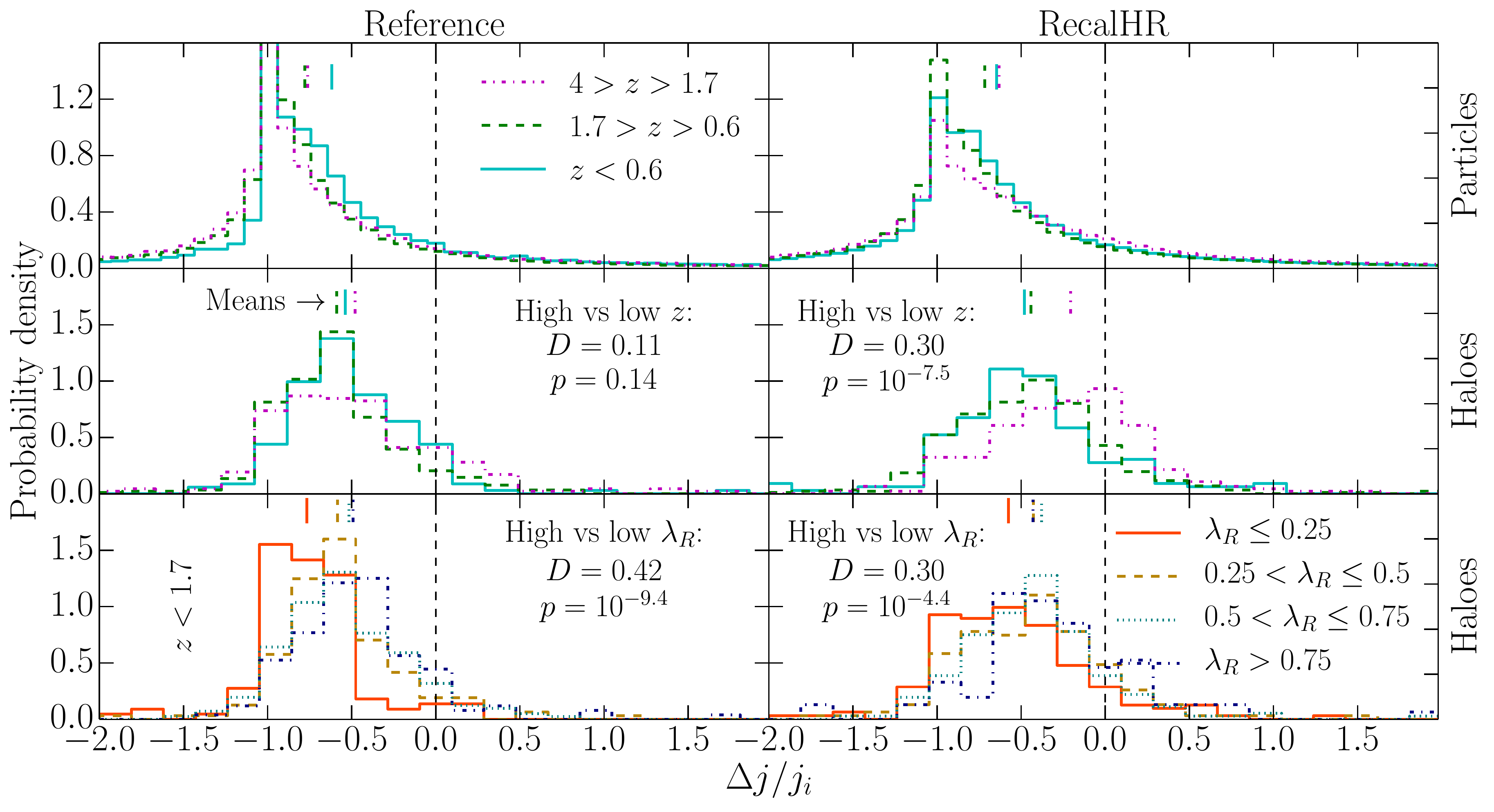}
	\caption{Relative change in the specific angular momentum component parallel to the rotation axis of cooling particles during cooling episodes, compared to the initial state.  The left-hand panels include haloes from our Reference EAGLE sample, while the right-hand panels include those from the RecalHR run.  The top panels consider particles individually.  The middle and bottom panels consider the change in net specific angular momentum of all cooling particles in a single episode, i.e.~on a halo-by-halo basis.  Where the top two rows of panels bin by redshift, the bottom row of panels bins by the $\lambda_R$ value of the central galaxy for all $z<1.7$. The short, vertical marks give the mean of each distribution for respective linestyles.  If angular momentum were always conserved along the cooling axis, the distributions would be $\delta$ functions matching the vertical, dashed line.  Instead, we see angular-momentum losses in the cooling gas.  We perform Kolmogorov--Smirnov tests to compare the consistency of the halo distributions between $4>z>1.7$ and $z<0.6$, and between the distributions for $\lambda_R \leq 0.25$ and $\lambda_R > 0.75$.  The $D$-value printed in the panels gives the maximum vertical separation of the cumulative probability distributions, while the $p$-value is the probability that the means of the distributions would be as separated as they are under the pretext that they come from the same underlying population.  A little less angular momentum appears to be lost when gas cools onto a rotationally supported galaxy.} 
	\label{fig:Delta_j}
\end{figure*}

We measure the specific angular momentum, $j$, of cooling and cooled particles (i.e.~immediately before and after cooling) in our EAGLE haloes along their respective \emph{cooling} particles' axes of rotation on an individual basis and for the summed quantity of all cooling particles involved in a single episode.  The relative change in $j$ for each case is presented in the upper and middle panels of Fig.~\ref{fig:Delta_j}, respectively.  These two cases allow us to distinguish between `strong' and `weak' angular-momentum conservation \citep{fall02}.  In the `strong' case, individual particles would conserve $j$, whereas only the net $j$ of a collection of particles needs to be conserved in the `weak' case.  We bin our systems by redshift, as in the previous sections, and show distributions for the relative change in angular momentum for each of these bins.  Because resolution is known to play a role in $j$ losses, we present both the Reference and RecalHR simulations in the left and right panels of Fig.~\ref{fig:Delta_j}, respectively.

It is clear from the top panels of Fig.~\ref{fig:Delta_j} that individual particles do not conserve angular momentum while cooling, and thus EAGLE haloes do not demonstrate strong $j$ conservation.  Comparing this to the middle panels, those particles appear to be exchanging some of their angular momentum between one another, as the net change in $j$ for the collection of particles provides a narrower distribution with a peak closer to zero.  Yet, in both cases, the means of the distributions are negative.  Therefore, on average, \emph{angular momentum is lost by gas as it cools} onto EAGLE galaxies; i.e.~even weak $j$ conservation is not satisfied.  There is no clear evolution in the distributions for either the upper or middle panels of Fig.~\ref{fig:Delta_j} for the Reference simulation; for the highest- to lowest-redshift bins, the average net loss of $j$ during a cooling episode is approximately 55, 64, and 57 per cent, respectively.  The highest- and lowest-redshift histograms are also consistent according to the Kolmogorov--Smirnov test (see Fig.~\ref{fig:Delta_j}).  We find that at higher resolution (the RecalHR run), $j$ losses during cooling are reduced at higher redshift, on average.  We suggest that, by itself, this is not enough to claim any generic correlation between $\Delta j / j_i$ and $z$, especially as the results between simulations agree at low redshift.  We come back to this point in Section \ref{ssec:relj}.

Galactic discs in EAGLE are known to be realistic in their size \citep{schaye15,lange16,furlong16}.  Yet, the gas that cools loses a fraction of angular momentum that is consistent with the large percentages reported when the `angular-momentum catastrophe' produced simulated discs that were too concentrated \citep[cf.][]{katz91}.  While, indeed, too much \emph{absolute} angular momentum was lost during cooling in early hydrodynamic, cosmological simulations, evidently the correct amount of \emph{specific} angular momentum was lost, based on our results.

The frequency of particle interactions and collisions will determine the potential for angular-momentum loss of cooling particles.  We hypothesise that in a rotationally supported system, where there is less random motion, collisions might be fewer, and thus less angular momentum might be lost.  To test this, we first quantify the relative level of rotation and dispersion support in our galaxies using the $\lambda_R$ parameter, introduced by \citet{emsellem07}.  Galaxies with $\lambda_R \sim 0$ are predominantly dispersion-supported, whereas those with $\lambda_R \sim 1$ are predominantly rotationally supported.  For our EAGLE galaxies, we measure the property as
\begin{equation}
\lambda_R = \frac{\sum_* m_*\, j_{z,*}}{\sum_* m_* \sqrt{j_{z,*}^2 + r_*^2\, v_{z,*}^2}}~,
\end{equation}
where the sums are over all star particles associated with the main subhalo within the `BaryMP' galactic radius defined by \citet[][where the cumulative mass profile of the stars and cold gas reaches a constant gradient]{stevens14}, and $j_z$ and $v_z$ are the specific angular momentum and velocity components along the galaxy's rotation axis, respectively, as measured in the galaxy's centre-of-momentum frame.\footnote{Note that this calculation for $\lambda_R$ does not include the intricacies required to compare against observations as in \citet{naab14}, as we simply require a relative measure of this quantity for internally comparing EAGLE galaxies.}  In the bottom panels of Fig.~\ref{fig:Delta_j}, we again show the relative net loss of specific angular momentum during cooling events, but now bin galaxies by $\lambda_R$, including all systems for $z<1.7$.  By excluding those in the range $1.7<z<4$, we eliminate the population of high-$z$ systems from RecalHR that we have shown lose less $j$ during cooling.  We would have otherwise had biased results, as the average $\lambda_R$ of the galaxies is lower at higher $z$.  

For the Reference and RecalHR simulations, we find a weak tendency for $j$ losses to be reduced for high-$\lambda_R$ systems.  To quantify this statistically, we have compared the distributions for the highest and lowest $\lambda_R$ histograms in the bottom panels of Fig.~\ref{fig:Delta_j} using the Kolmogorov--Smirnov test, the results of which are given in the panels.  The low $p$-values ($10^{-9.4}$ and $10^{-4.4}$ for the Reference and RecalHR simulations, respectively) suggest there is a non-negligible statistical significance to the separation of the means of the distributions, although this is less significant for the higher-resolution simulation.  We thus \emph{tentatively} find these results in favour of our hypothesis.  Our results suggest any reduction in $j$ losses during cooling caused by stronger rotation in the central galaxy is, at most, a few tens of per cent.

\subsection{Relative orientation and magnitude of specific angular momenta}
\label{ssec:relj}

As already discussed, galaxy formation models typically assume that the net specific angular momentum of hot gas about to cool, $\vec{j}_{\rm cooling}$, is equivalent to that of all the hot gas in the halo, $\vec{j}_{\rm hot}$, and to that of the entire halo itself, $\vec{j}_{\rm halo}$, both in terms of magnitude and direction.  Without the ability to directly measure the motion of dark matter in haloes to independently measure a halo's spin, it is impossible to determine if $\vec{j}_{\rm cooling}$ and $\vec{j}_{\rm halo}$ are the same empirically with observational methods.  Simulations are the only current means of addressing this in any capacity.  Previous studies of cosmological, hydrodynamic simulations have shown that gas and dark matter in haloes tend to have different and/or offset specific angular momenta \citep*[e.g.][]{bosch02,chen03,bosch03,sharma05,sharma12}.  Attention has not been given specifically to the \emph{cooling} particles before, however.

\subsubsection{Magnitude}
\begin{figure*}
	\centering
	\includegraphics[width=0.754\textwidth]{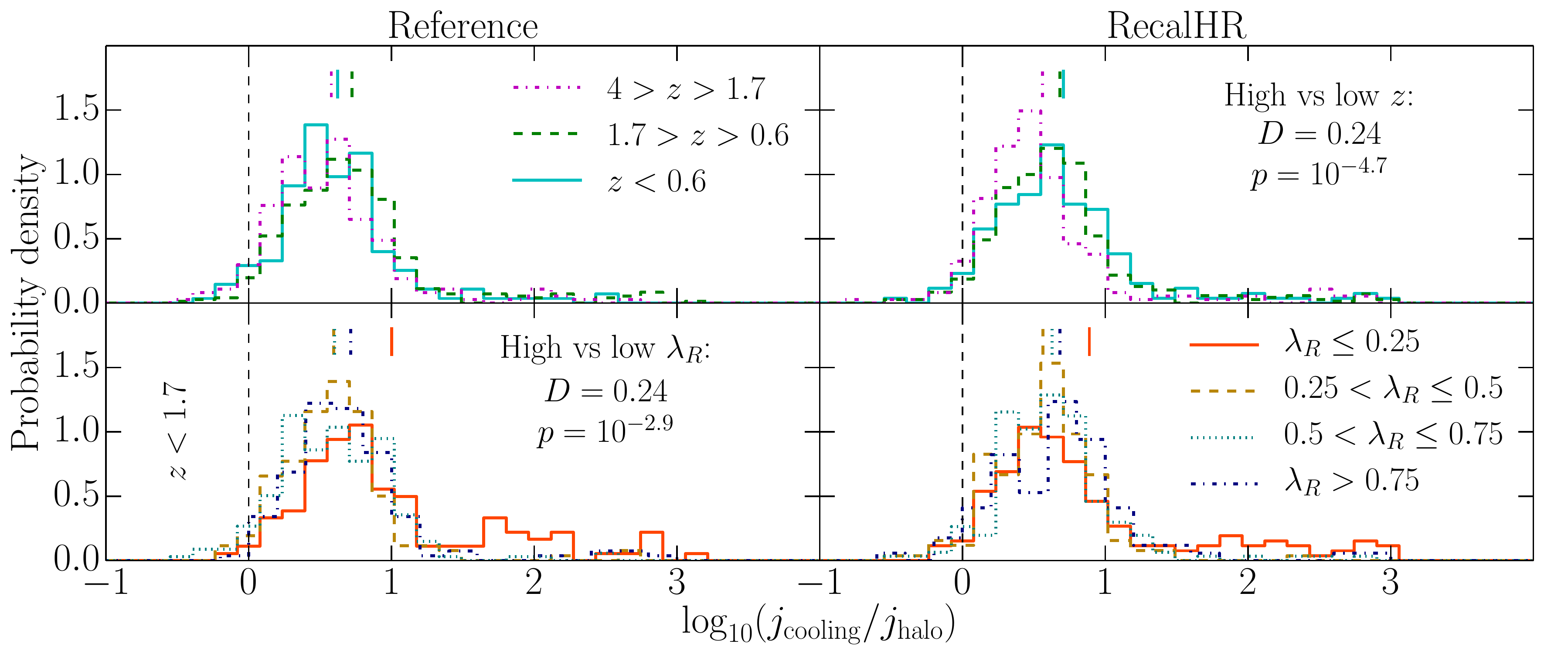}
	\caption{Probability distributions for the ratio of the magnitudes of net specific angular momentum of cooling gas to that of the halo, drawn from our sample of Reference and RecalHR EAGLE systems.  The top panels cover all systems in each redshift bin, while the bottom panels cover all $z<1.7$ and bins by $\lambda_R$ of the central galaxy.  The dashed, vertical line indicates where $j_{\rm cooling} = j_{\rm halo}$, which is often assumed in galaxy formation models.  The short, vertical marks give the means of the distributions.  We perform Kolmogorov--Smirnov tests on the consistency of the distributions in each panel, and note the cases where there is potential inconsistency (see caption of Fig.~\ref{fig:Delta_j}).  Cooling gas carries more specific angular momentum than the halo on almost all occasions.}
	\label{fig:j_cool_halo}
\end{figure*}

We directly measure the ratio $j_{\rm cooling}/j_{\rm halo}$ from our EAGLE haloes, and present distributions for this ratio for bins of redshift in the top panels of Fig.~\ref{fig:j_cool_halo}.  We see that cooling gas typically has more specific angular momentum than the halo.  This result is contrasting (but not opposing) to the combined results of observations and models that suggest the stellar content of galaxies typically has lower specific angular momentum than their haloes \citep[e.g.][]{rf12,stevens16}.  As discussed in Section \ref{ssec:cons}, some of this angular momentum is lost during the cooling process.  As seen in the top panels of Fig.~\ref{fig:j_cool_halo}, there is no \emph{clear} evolution in $j_{\rm cooling}/j_{\rm halo}$ for our EAGLE haloes.  We do, however, find the mean ratio to be lower for RecalHR at higher redshift, which appears to be statistically significant, based on a Kolmogorov--Smirnov test (see the top right panel of Fig.~\ref{fig:j_cool_halo}).  This is balanced by the fact that less angular momentum is lost during the cooling process for these specific haloes, as shown in Section \ref{ssec:cons}.  If we compare these findings with the lower left panel of Fig.~\ref{fig:tempref}, we see that the temperature of the cooling gas in these haloes is lower at the `beginning' of the cooling episode for the RecalHR run.  The evidence suggests the population of RecalHR haloes in our sample for $4>z>1.7$ effectively had a head start in cooling over the same haloes in the Reference simulation, and thus were measured at a moment when the gas was already cooler and had already lost some of its specific angular momentum.  This would imply the angular momentum measurements in the Reference and RecalHR runs are entirely consistent.

We have already shown tentative evidence that gas cooling onto a rotationally supported galaxy in EAGLE loses less specific angular momentum than that cooling onto a dispersion-supported galaxy (Fig.~\ref{fig:Delta_j}).  In the bottom panels of Fig.~\ref{fig:j_cool_halo} we also break $j_{\rm cooling}/j_{\rm halo}$ into bins of $\lambda_R$.  There is a population of low-$\lambda_R$ galaxies, in both the Reference and RecalHR simulations, that show high values for $j_{\rm cooling}/j_{\rm halo}$, but are not abnormal in any other respect we can find.  We find nearly all the excess specific angular momentum is lost during cooling for these systems.  Modulo those few systems, the distributions for $j_{\rm cooling}/j_{\rm halo}$ for various $\lambda_R$ are entirely consistent with each other.  In other words, gas should not be aware of what type of galaxy it will cool onto at the moment it begins to cool, which appears to indeed be the case for EAGLE.

\begin{figure}
	\centering
	\includegraphics[width=0.96\textwidth]{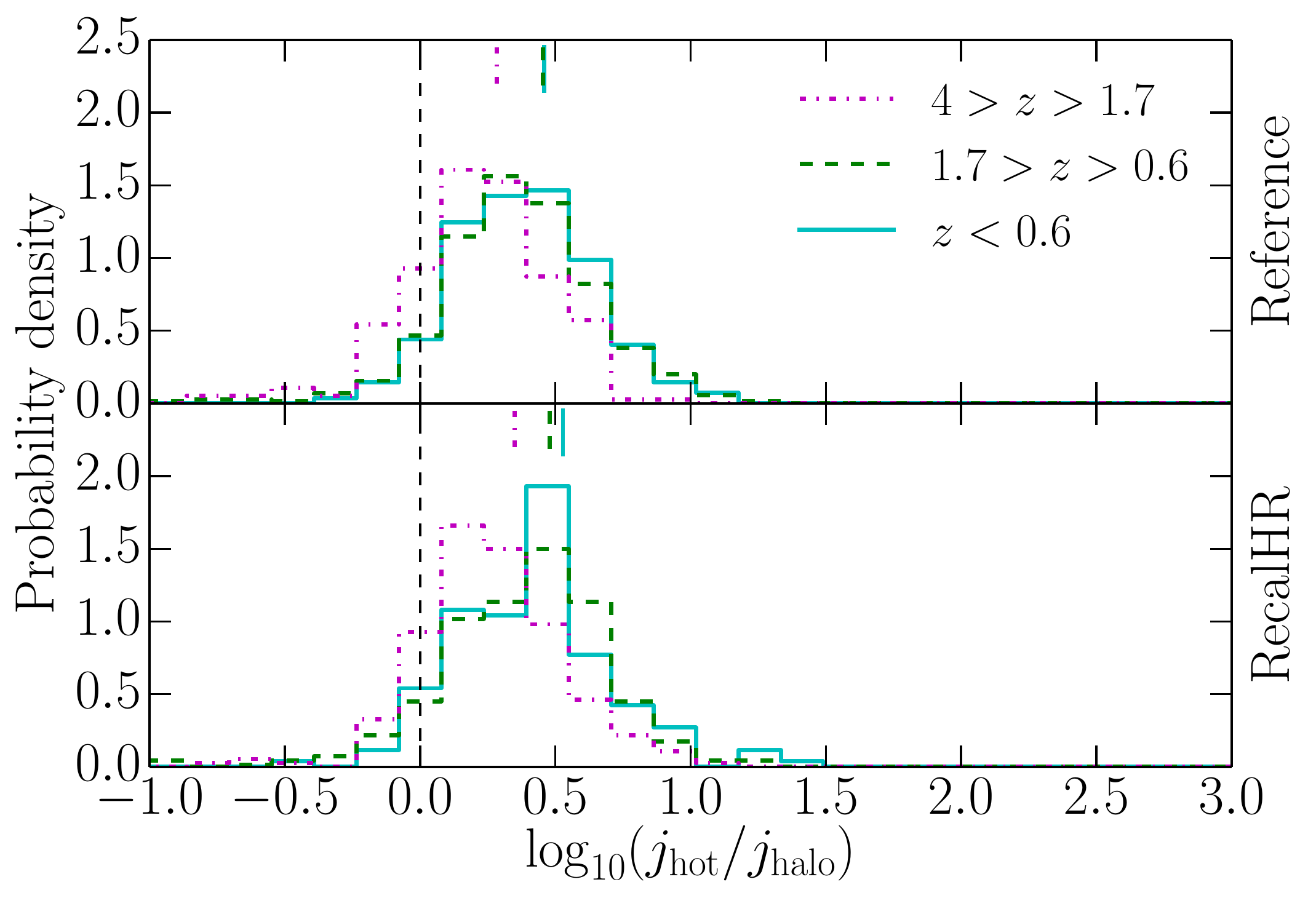}
	\caption{Normalised histograms of the ratio of the magnitudes of net specific angular momentum of hot gas to that of the entire halo (within $R_{\rm vir}$) for our Reference (top panel) and RecalHR (bottom panel) EAGLE systems for three redshift bins.  The vertical, dashed line indicates where $j_{\rm hot} = j_{\rm halo}$.  The short, vertical marks are the means for each distribution.  Hot gas in haloes has preferentially higher specific angular momentum than the halo as a whole.}
	\label{fig:j_hot_halo}
\end{figure}

\begin{figure}
	\centering
	\includegraphics[width=0.96\textwidth]{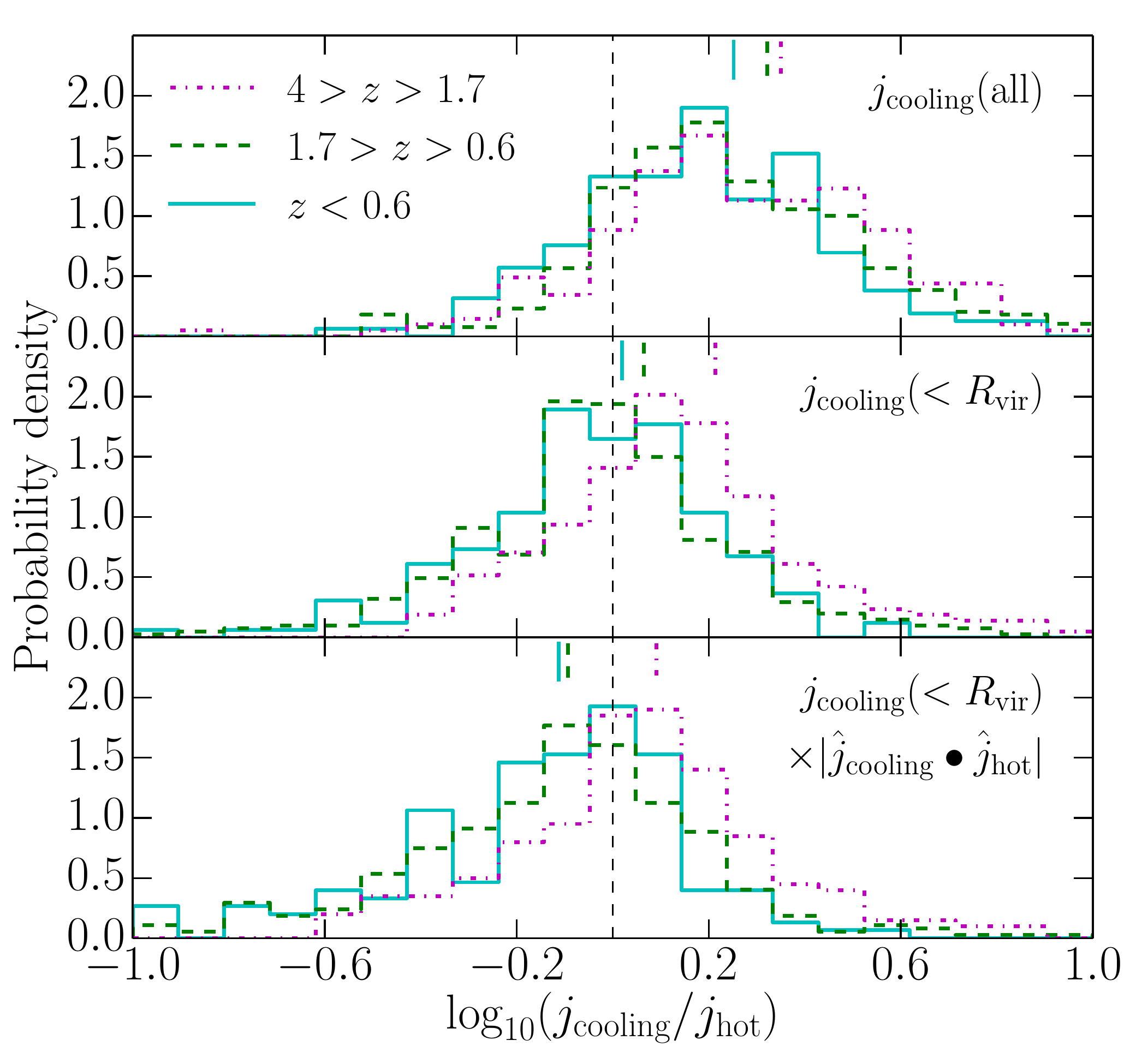}
	\caption{Normalised histograms of the ratio of the magnitudes of net specific angular momentum of cooling gas to that of hot gas for our Reference EAGLE systems.  In all panels, $j_{\rm hot}$ is calculated using only the particles within the virial radius.  The top panel calculates $j_{\rm cooling}$ using all cooling particles.  The middle panel only considers cooling particles within the virial radius.  The bottom panel presents the $j_{\rm cooling}$ component for particles within the virial radius projected onto the axis of rotation of the hot particles.  The vertical, dashed line indicates where $j_{\rm cooling} = j_{\rm hot}$.  The short, vertical marks are the means for each distribution.}
	\label{fig:j_cool_hot}
\end{figure}

It is not just the cooling gas whose $j$ exceeds that of the halo.  In fact, hot gas in general has preferentially high $j$ in our EAGLE haloes, as shown in Fig.~\ref{fig:j_hot_halo}.  This result is consistent with previous studies of non-radiative hydrodynamic simulations \citep{chen03,sharma05}, but contends with the earlier work of \citet{bosch02}.  The fact that $\langle j_{\rm hot}/j_{\rm halo}\rangle >1$ automatically explains why $\langle j_{\rm cooling}/j_{\rm halo}\rangle >1$.  Yet, we also find that $\langle j_{\rm cooling}/j_{\rm hot}\rangle >1$ for our haloes.  This can be seen by comparing Figs.~\ref{fig:j_cool_halo} \& \ref{fig:j_hot_halo} and is shown more explicitly in the top panel of Fig.~\ref{fig:j_cool_hot}.  A contributing factor to this is that $j_{\rm hot}$ has been measured exclusively for hot particles within $R_{\rm vir}$, while $j_{\rm cooling}$ considers \emph{all} cooling particles here.  Once the virial-radius restriction is imposed on $j_{\rm cooling}$ too, the highest-$j$ particles are eliminated, and so this becomes more consistent with $j_{\rm hot}$, as seen in the middle panel of Fig.~\ref{fig:j_cool_hot}.  Still, when we average over our full halo sample at all redshifts, we find $\langle j_{\rm cooling}(<R_{\rm vir})/j_{\rm halo} \rangle \simeq 1.4$.  Most of this extra $j_{\rm cooling}$ comes from a non-zero component orthogonal to $\hat{j}_{\rm hot}$.  This is demonstrated by the bottom panel of Fig.~\ref{fig:j_cool_hot}, which now projects $\vec{j}_{\rm cooling}$ onto $\hat{j}_{\rm hot}$.  Here, the projected magnitude of $j_{\rm cooling}$ is consistent with $j_{\rm hot}$, once averaged over all redshifts.

If one assumes that gas particles at the centre of the halo have lower $j$ than those at the outskirts, then our result from Fig.~\ref{fig:rhohot} (that cooling gas originates preferentially from the halo centre, regardless of the underlying hot-gas distribution, while the hot gas forms a core) can explain why there is an evolutionary decline in $j_{\rm cooling}/j_{\rm hot}$, seen in all panels of Fig.~\ref{fig:j_cool_hot}.  We remind the reader that because we looked for particles that transitioned from hot to cold over time-steps of several hundred Myr, we will have missed any particles that both cooled and were reheated over that time.  While we suggested in Section \ref{ssec:sample} that the fraction of these particles is generally small, the lowest-$j$ particles are preferentially reheated by feedback \citep[see][]{brook11,brook12}, so our $j_{\rm cooling}$ values are more like close upper limits.  Thankfully, $j_{\rm hot}$ is not subject to this bias, so our finding that gas has specific angular momentum in excess of $j_{\rm halo}$ when it begins to cool in EAGLE haloes is robust.

\subsubsection{Orientation}
\begin{figure*}
	\centering
	\includegraphics[width=0.754\textwidth]{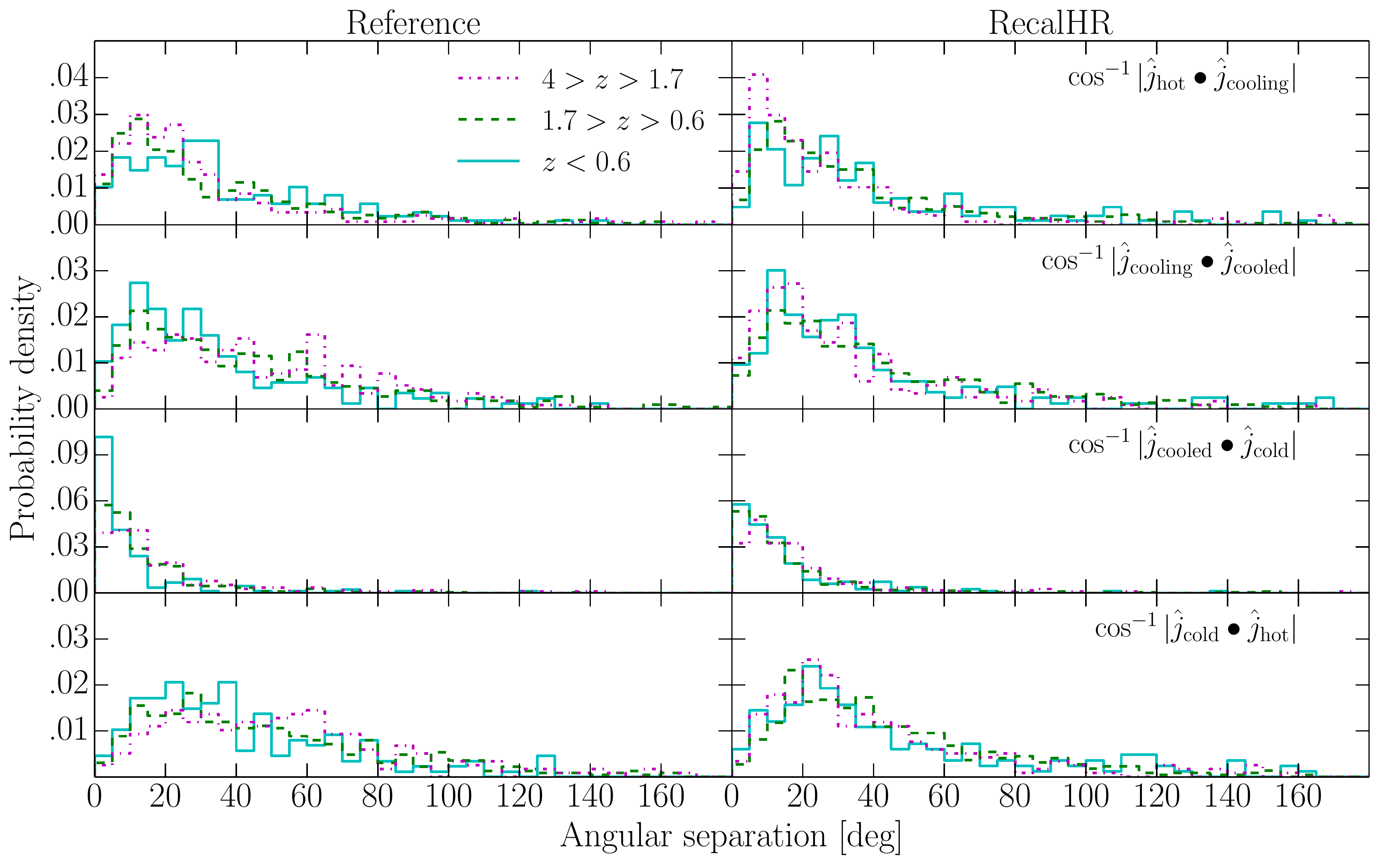}
	\caption{Angular separations between rotation axes of gas particle groups during cooling.  The top panels give the difference between all the hot particles and those about to cool.  The second row of panels is the change in the rotation axis of the particles that cool over the cooling episode.  The third row of panels are for all cold gas and the particles that have just cooled.  The bottom panels compare the rotation axes of all cold gas to all hot gas.  Cooling gas need not be aligned with the rest of the hot gas in the halo, and it precesses during the cooling process to become aligned with gas that is already cold.}
	\label{fig:offsets}
\end{figure*}

The direction of the angular momentum of the cooling gas is also important for how the new material will alter the disc.  In the top panels of Fig.~\ref{fig:offsets} we show the angular offsets of the spin vectors of hot gas about to cool and hot gas in general for our EAGLE haloes.  Regardless of redshift, these are typically offset by tens of degrees.  The second row of panels in Fig.~\ref{fig:offsets} shows the angular difference in rotation axis of the cooling and cooled particles; that is before and after the transition from hot to cold.  Again, differences of tens of degrees are common, indicating that gas particles precess during cooling.  In fact, they precess to come in line with the pre-existing cold gas in the galaxy, as seen by the small angular differences in the rotation axes of the cooled and cold particles in the third row of panels in Fig.~\ref{fig:offsets}.  A combination of the cooling gas's rotation axis being consistently offset from the hot gas as a whole with the gas being subject to precession sets up an offset between the rotation axes of the cold gas in the galaxy and its hot halo.  This is shown by the bottom panels of Fig.~\ref{fig:offsets}, where, once again, offsets of many tens of degrees are common and persist to $z=0$.  Having such an offset can lead to warped discs \citep{roskar10}.  As with our other results, we do not find a clear, systematic difference between the Reference (left panels of Fig.~\ref{fig:offsets}) and RecalHR (right panels) runs, suggesting this is indeed a physical result.

\begin{figure}
	\centering
	\includegraphics[width=0.95\textwidth]{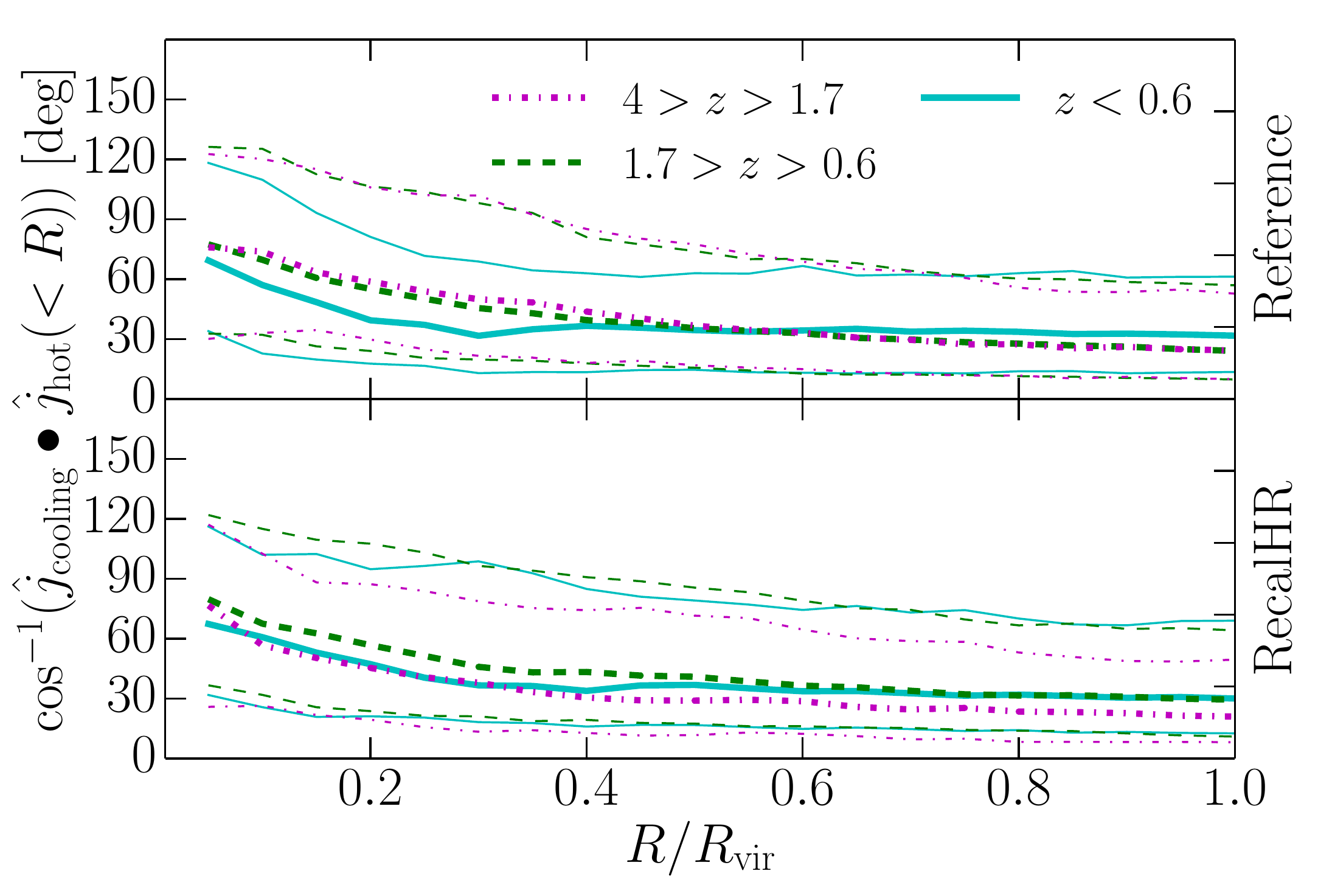}
	\caption{Angular separation between spin directions of hot gas in haloes within a given radius and all cooling gas for the same haloes.  Cooling gas is poorly aligned with hot gas on all scales, regardless of redshift or resolution.}
	\label{fig:offsets_r}
\end{figure}

Because most cooling occurs in the inner part of the hot halo (Fig.~\ref{fig:rhohot}), one might also expect the inner hot halo's rotation axis to be better aligned with the cooling gas.  We calculate whether the rotation axis offset between cooling and hot gas varies when only hot gas internal to a given radius is included, which we present in Fig.~\ref{fig:offsets_r}.  Contrary to expectation, we find the inner hot gas is typically \emph{less} aligned with the cooling gas.  Any evolution in the profiles in Fig.~\ref{fig:offsets_r} is minimal for our sample of EAGLE haloes, as seen by the similarity of the curves in Fig.~\ref{fig:offsets_r}.  The Reference (upper panel) and RecalHR (lower panel) simulations again produce the same result, so we can trust that this is not predominantly an effect of resolution.  Our result is contrasting (but not at all opposing) to the results of \citet{zavala16}, who find that the evolution of $j$ of stars in EAGLE galaxies is more closely tied to that of the inner dark-matter halo ($< 0.1 R_{\rm vir}$), rather than the halo in its entirety.

The relative orientations and magnitudes of $j$ for various particle types in haloes \emph{in general} is an interesting topic which we are investigating in detail in an accompanying paper (Contreras et al.~in preparation).  The sample of cooling systems used here is consistent with the differences in direction and magnitude of $j$ between baryons and dark matter in all EAGLE haloes to be presented there.  That paper will address uncertainties in angular momentum as a function of the number of particles used; in general $\gtrsim$100 particles is required for trustworthy measurements of individual haloes.  We do not find any significant changes to our conclusions here (which are concerned with the \emph{population}) if we exclude haloes with $<$100 particles though.

\section{After cooling: cold gas in galaxies}
\label{sec:cold}
\subsection{Radial surface density profiles of gas discs}
\label{ssec:Sigma_r}

In the picture of disc formation proposed by \citet{fall80}, gas cools and collapses to form an exponential disc while conserving angular momentum:
\begin{equation}
\Sigma_{\rm cooled}(r) = \Sigma_0 e^{-r/r_{\rm d}} \simeq \frac{m_{\rm cooled}}{2 \pi r_{\rm d}^2} e^{-r/r_{\rm d}}~,
\label{eq:Sigma_cool}
\end{equation}
where $r_{\rm d} \ll R_{\rm vir}$ is the scale radius of the disc.  The exponentiality of discs is a typical assumption of galaxy formation models, where some even maintain that a disc must \emph{always} be exponential \citep[e.g.][]{cole00,hatton03,croton06,sage,somerville08a,guo11}.  But only relatively recently have resolved observations of the H\,{\sc i} and CO (H$_2$) distribution in local spiral galaxies been made \citep[e.g.][]{walter08,leroy09}, allowing us to actually see what they are like. The 33 galaxies analysed by \citet{bigiel12} suggest that while an exponential profile broadly describes the average galaxy at intermediate radii, there is plenty of deviation from exponentials, and a strong suggestion of cusps existing at the centres of these discs.  With EAGLE, we can not only check the exponentiality of gas discs, but also compare directly to equation (\ref{eq:Sigma_cool}).

\begin{figure*}
	\centering
	\includegraphics[width=0.95\textwidth]{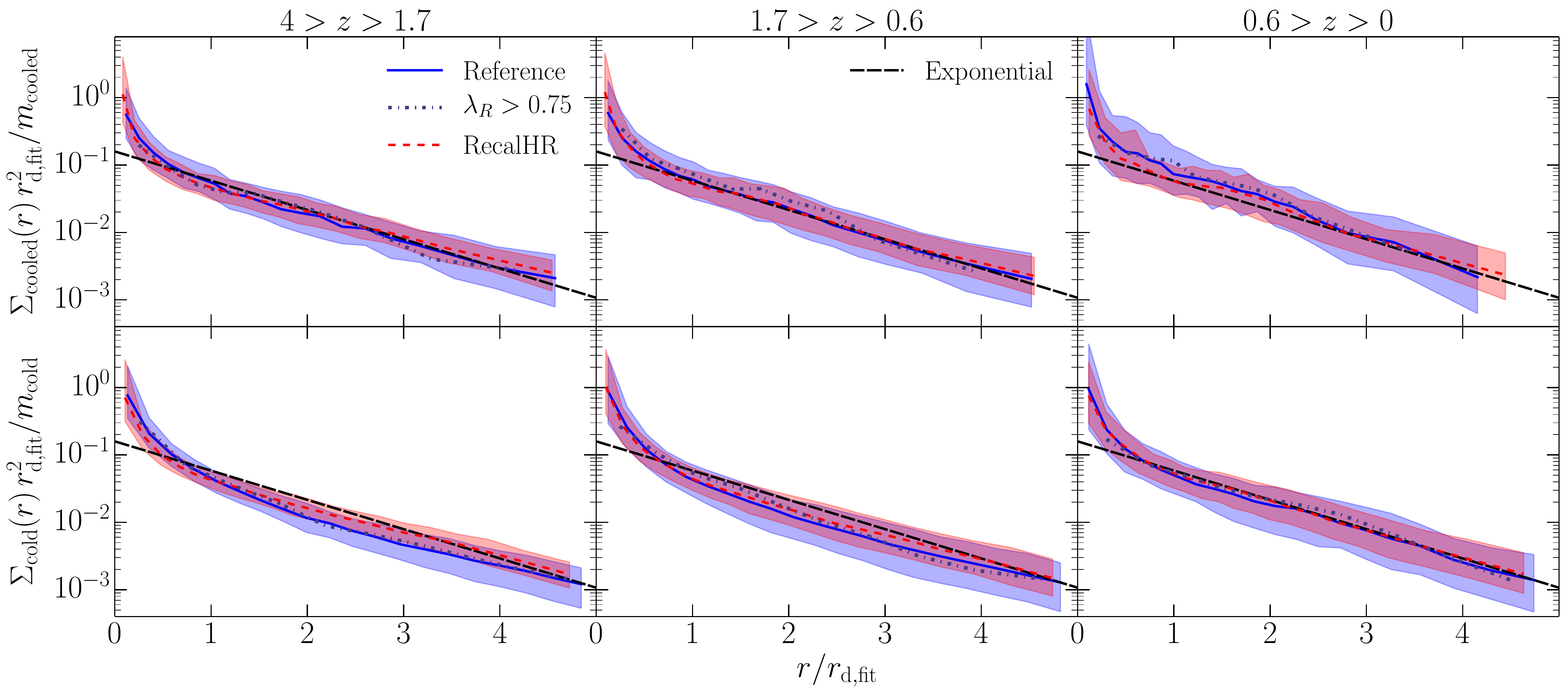}
	\caption{Surface density profiles for cold gas particles in our sample of EAGLE galaxies.  As for previous figures, we present 3 redshift ranges for the Reference and RecalHR simulations, presenting their median profiles (solid and short-dashed curves, respectively) and 16$^{\rm th}$--84$^{\rm th}$ percentile ranges (shaded regions).  Radial distances have been normalised for each galaxy by their fitted exponential scale radius, $r_{\rm d,fit}$.  The top row of panels considers only the recently cooled particles, whereas the bottom panels consider all cold particles.  All surface densities have been normalised by the amount of cooled/cold gas considered and the square of the scale radius.  The long-dashed line is a precise exponential (equation \ref{eq:Sigma_cool}).  The dot-dashed curves give the median profile for the rotation-dominated galaxies ($\lambda_R > 0.75$) for each subsample of the Reference haloes.  With the addition of a central cusp, an exponential describes the profile of recently cooled gas well, but only works as well for cold discs as a whole at low redshift.}
	\label{fig:Sigma_r}
\end{figure*}

We present surface density profiles for recently cooled and all cold gas of our sample of EAGLE haloes in Fig.~\ref{fig:Sigma_r} (top and bottom panels, respectively).  To build these profiles, we find the rotation axis of the cold (cooled) particles to determine the plane of the galaxy (freshly cooled gas), and measure the surface density in annuli\footnote{For our intents and purposes, an annulus is a cylindrical shell that cannot exceed the virial sphere in height.} in this plane.  Consistent with our measurements of the hot gas profiles, the widths of the annuli are adapted to contain the same number of particles, while respecting the softening scale.  100 annuli are used for the cold gas, and 20 are used for the cooled gas.  We then fit an exponential to each profile and normalise the annuli's radii by the fitted scale length.  By normalising the surface densities as well, we are able to see how well an exponential describes the entire galaxy population as a whole.  Note that $r_{\rm d,fit}$ is a different quantity for the $\Sigma_{\rm cooled}(r)$ and $\Sigma_{\rm cold}(r)$ profiles in Fig.~\ref{fig:Sigma_r}.

The top panels of Fig.~\ref{fig:Sigma_r} show that $\Sigma_{\rm cooled}(r)$ for EAGLE galaxies is nicely centred on equation (\ref{eq:Sigma_cool}), with a scatter increasing at lower redshift (the 68\% confidence range is 0.53, 0.62, and 0.79 dex tall on average for the highest- to lowest-redshift bins, respectively).  The profiles include a cusp, seen at $r \lesssim 0.6 r_{\rm d}$ for the highest-redshift bin, with the prominence of this feature growing modestly down to $z=0$.  Approximately 20 per cent of the cooled gas mass lies inside $r < 0.6 r_{\rm d}$ when averaged over all galaxies in our sample for either the Reference or RecalHR simulation.  If these profiles were exponential all the way to the centre, only 12 per cent of their mass should lie inside $r < 0.6 r_{\rm d}$.

It is worth highlighting the fact that, because every EAGLE profile has been normalised by a fitted exponential scale radius, if an exponential were a general, genuine, good fit for the surface density profiles in Fig.~\ref{fig:Sigma_r}, the EAGLE profiles should cluster about the long-dashed line with a small amount of scatter.  A cusp can still be recovered from an otherwise exponential profile, provided the number of points being fitted to in each cusp is not a large fraction of those in the total profile.  This is what is seen in the top panels.  On the other hand, the bottom panels of Fig.~\ref{fig:Sigma_r} show that, at higher redshift, the $\Sigma_{\rm cold}(r)$ profiles are curved for the entire plotted range (cf.~the long-dashed line).  This implies an exponential (even with a cusp) is too simplistic to describe these profiles in general.

Of course, the galaxies in our sample need not be all rotationally supported, and hence may not have well-behaving, classical discs.  To this point, we over-plot the median surface density profiles for cold and recently cooled gas for the rotation-dominated galaxies with $\lambda_R > 0.75$ in Fig.~\ref{fig:Sigma_r} with dot-dashed curves.  By selecting galaxies we expect to be more disc-like, we find no difference in the normalised surface density profiles for cold or recently cooled gas.  While the recently cooled gas in these galaxies lost marginally less $j$ during the cooling process (Fig.~\ref{fig:Delta_j}), this has not translated into a difference in $\Sigma_{\rm cooled}(r)$.

In a semi-analytic model, one must put in what the scale length of the cooling disc is.  If one assumes gas cools into an exponential disc, that $ j_{\rm cooled} = j_{\rm halo}$, and that the disc has a flat rotation velocity with $v_{\rm rot} = V_{\rm vir}$, it is straight forward to show that the scale radius of that exponential disc is
\begin{equation}
r_{\rm d,model} \equiv \frac{j_{\rm halo}}{2 V_{\rm vir}}
\label{eq:r_d}
\end{equation}
\citep[e.g.][]{fall83,mo98}.  Equation (\ref{eq:r_d}) has formed the basis of determining (initial) disc sizes in many semi-analytic models \citep[e.g.][]{hatton03,croton06,fu10}, but more complex algorithms are also popular \citep[e.g.][]{cole00}.  

We showed in Section \ref{sec:angmom} that the assumptions about the angular momentum of the cooling gas required for equation (\ref{eq:r_d}) to be accurate do not agree with EAGLE, so we would expect some difference between the model scale length, $r_{\rm d,model}$, and the fitted scale length, $r_{\rm d,fit}$.  After measuring the net specific angular momentum of each halo in our EAGLE sample as
\begin{equation}
j_{\rm halo} = \frac{| \sum_p m_p \vec{r}_p \times \vec{v}_p |}{\sum_p m_p}
\label{eq:j_halo}
\end{equation}
(where subscript $p$ is for particles of all types within $R_{\rm vir}$), we can quantify how discrepant $r_{\rm d,model}$ is from $r_{\rm d,fit}$.

\begin{figure}
	\centering
	\includegraphics[width=0.95\textwidth]{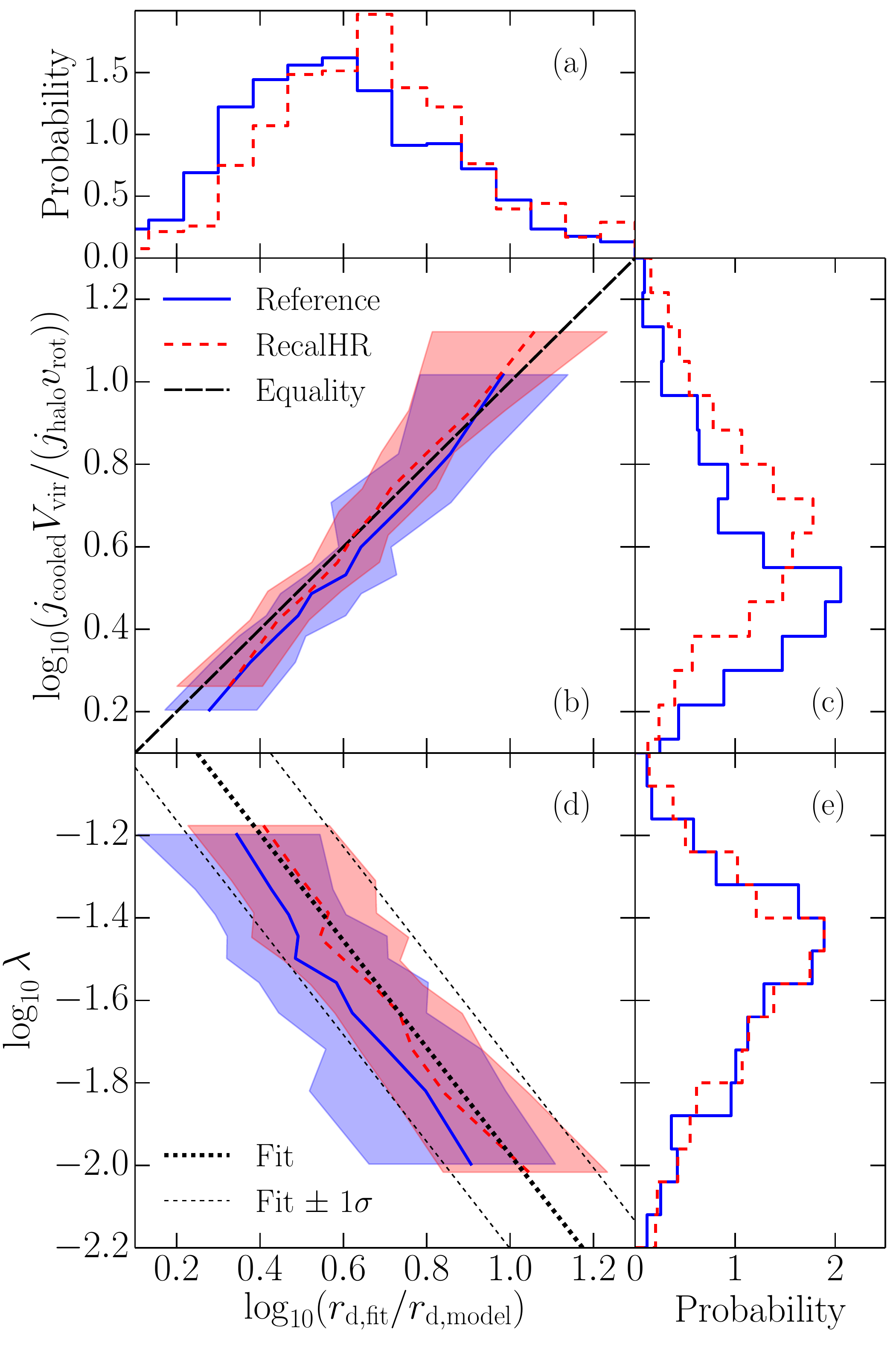}
	\caption{Trends relating to the ratio of the fitted exponential scale length of the surface density profiles of recently cooled gas in EAGLE galaxies to a typical assumed scale length in semi-analytic models (equation \ref{eq:r_d}).  \emph{Panel (a):} Normalised histograms for this ratio for each of the Reference and RecalHR simulations. \emph{Panel (b):} Relation between this ratio and a combination of the specific angular momentum of the cooled gas, that of the halo, the virial velocity of the halo, and the rotational velocity of the cooled gas.  These quantities are expected to be nearly equal (equation \ref{eq:rsrd}, as given by the long-dashed line).  The solid and short-dashed curves give the median for the Reference and RecalHR simulations respectively, with the 16$^{\rm th}$--84$^{\rm th}$ percentile range for each shown by the shaded regions (for data binned along the $y$-axis). \emph{Panel (c):} Normalised histograms for the $y$-axis value of panel (b).  \emph{Panel (d):} Relation between the scale radius ratio and the spin parameter of the halo.  In addition to the median and 16$^{\rm th}$--84$^{\rm th}$ percentile range (for data binned along the $y$-axis), we include a least-squares fit for the RecalHR simulation, including the standard deviation for points about this fit, where $\log_{10} \lambda$ has been taken as the independent variable.  \emph{Panel (e):} Normalised histograms for the spin parameter of the haloes.  Only in the highest-spin haloes do the specific angular momenta of the cooled gas and halo become similar enough for the scale radius to resemble the model value.}
	\label{fig:rsrd}
\end{figure}

In Fig.~\ref{fig:rsrd}(a), we present normalised histograms for the ratio of $r_{\rm d,fit}/r_{\rm d,model}$ for both the Reference and RecalHR haloes, for all $z<4$.  Both simulations find an almost log-normal distribution for this ratio, where it is almost always true that $r_{\rm d,fit} > r_{\rm d,model}$.  The idea that $r_{\rm d} \propto j_{\rm halo}/V_{\rm vir}$ (equation \ref{eq:r_d}) came from the assumptions that $j_{\rm cooled} = j_{\rm halo}$ and that the cooled gas all rotated with velocity $V_{\rm vir}$.  In practice, we should really expect $r_{\rm d} \approxprop j_{\rm cooled}/v_{\rm rot}$, where $v_{\rm rot}$ is mass-weighted mean tangential velocity of the cooled gas.  This then gives
\begin{equation}
\frac{r_{\rm d,fit}}{r_{\rm d,model}} \simeq \frac{j_{\rm cooled} V_{\rm vir}}{j_{\rm halo} v_{\rm rot}}~.
\label{eq:rsrd}
\end{equation}

In Fig.~\ref{fig:rsrd}(b), we compare the left- and right-hand side of equation (\ref{eq:rsrd}) for the Reference and RecalHR simulation haloes, displaying the median and 16$^{\rm th}$--84$^{\rm th}$ percentile range for each.  Note that we have binned data along the $y$-axis of this plot (and Fig.~\ref{fig:rsrd}d, but all other plots in this paper bin along the $x$-axis).  Both simulations corroborate equation (\ref{eq:rsrd}), with the median trend for the RecalHR simulation (with $\gtrsim$512 cooled particles per halo) nearly matching it perfectly.  In addition to an exponential not describing \emph{every individual} cooling profile, the scatter in Fig.~\ref{fig:rsrd}(b) can be attributed to the cooling particles having an angular-momentum structure (see Section \ref{ssec:Sigma_j}) where the specific angular momentum of individual particles does not have a precise monotonic relationship with either position or rotational velocity.

It is useful to relate $r_{\rm d,fit} / r_{\rm d,model}$ to global properties of the halo, as, if there is a strong correlation with one, the prescription for calculating $r_{\rm d}$ in a semi-analytic model could be easily modified to better match the results of EAGLE, rather than assuming weak $j$ conservation.  We find that the halo property that best correlates with $r_{\rm d,fit} / r_{\rm d,model}$ is the spin parameter,
\begin{equation}
\lambda \equiv \frac{J |E|^{1/2}}{G M_{\rm vir}^{5/2}}~,
\end{equation}
where $J$ and $E$ are the total angular momentum and energy (kinetic + potential) of the halo, respectively \citep{peebles69}.  Note that we approximate the spin parameter as
\begin{equation}
\lambda = \frac{j_{\rm halo}}{\sqrt{2} V_{\rm vir} R_{\rm vir}}
\end{equation}
\`{a} la \citet{bullock01}, which would be precise if the total halo density distribution were a singular isothermal sphere.  This deviates from a $\lambda$ value obtained assuming an NFW profile \citep*{nfw96} by $\lesssim 10$\% for concentrations of $c \lesssim 7$ \citep[see][]{mo98}.  For our haloes, $\lambda$ is correlated more strongly with $r_{\rm d,fit} / r_{\rm d,model}$ than any of $R_{\rm vir}$, $V_{\rm vir}$, $M_{\rm vir}$, $j_{\rm halo}$, or $r_{\rm d,model}$ independently, with a Pearson correlation coefficient of $-0.60$ and a corresponding $p$-value (probability of obtaining that coefficient by chance) of $\sim$$10^{-84}$.

The relation between $\lambda$ and $r_{\rm d,fit} / r_{\rm d,model}$ is shown in Fig.~\ref{fig:rsrd}(d).  By performing a least-squares fit for $\log_{10}(r_{\rm d,fit} / r_{\rm d,model})$ as a function of $\log_{10} \lambda$ for the RecalHR simulation, we obtain the relation
\begin{equation}
\log_{10}\left(\frac{r_{\rm d,fit}}{r_{\rm d,model}}\right) = -0.77 \log_{10} \lambda - (0.52 \pm 0.18)~,
\end{equation}
where the uncertainty is the standard deviation of the points about the fit.  Combining this result and equation (\ref{eq:r_d}), we suggest a modified model for the scale radius of cooling gas in a halo of given size and spin:
\begin{equation}
\log_{10}\left(\frac{r_{\rm d}}{R_{\rm vir}}\right) = 0.23 \log_{10}\lambda - 0.67 (\pm 0.18)~.
\label{eq:rdnew}
\end{equation}
The same fit can be made for the Reference simulation; in that case, the model scale radius would be
\begin{equation}
\log_{10}\left(\frac{r_{\rm d}}{R_{\rm vir}}\right) = 0.30 \log_{10}\lambda - 0.65 (\pm 0.22)~.
\label{eq:rdnew_ref}
\end{equation}
Any of these equations can be applied directly in a semi-analytic model.  Although, note that this assumes $\lambda$ measured from a dark-matter-only simulation is equivalent to considering all matter in a halo in a hydrodynamic simulation (which might not be true -- \citealt{schaller15a}, for example, have shown that the masses of haloes in the EAGLE simulations are systematically higher when run without baryonic physics).  

In Fig.~\ref{fig:rsrd}(e), we show that our EAGLE haloes (both Reference and RecalHR) are representative in terms of their distribution of spins, which should be roughly log-normal \citep[cf.][]{barnes87,bullock01}.

\subsection{Angular-momentum structure of gas discs}
\label{ssec:Sigma_j}
There is a building trend for models of disc evolution to calculate local processes in annuli of specific angular momentum, rather than radius \citep[e.g.][]{stringer07,dutton09,stevens16}.  These have been partially motivated by the fact that galaxies have dynamic and non-uniform velocity structures.  One can rewrite equation (\ref{eq:r_d}) as an exponential function of specific angular momentum, thereby relaxing the requirement that cooling gas should settle with a constant rotational velocity:
\begin{equation}
\Sigma_{\rm cooled}(j) = \Sigma_0(j_{\rm d}) e^{-j / j_{\rm d}} 
\label{eq:Sigma_j}
\end{equation}
(e.g.~as implemented in the \textsc{Dark Sage} semi-analytic model; \citealt{stevens16}).  Here, we investigate the angular-momentum structure of the cooled and cold gas profiles of our EAGLE galaxies and determine whether equation (\ref{eq:Sigma_j}) can approximate them.

\begin{figure*}
	\centering
	\includegraphics[width=0.95\textwidth]{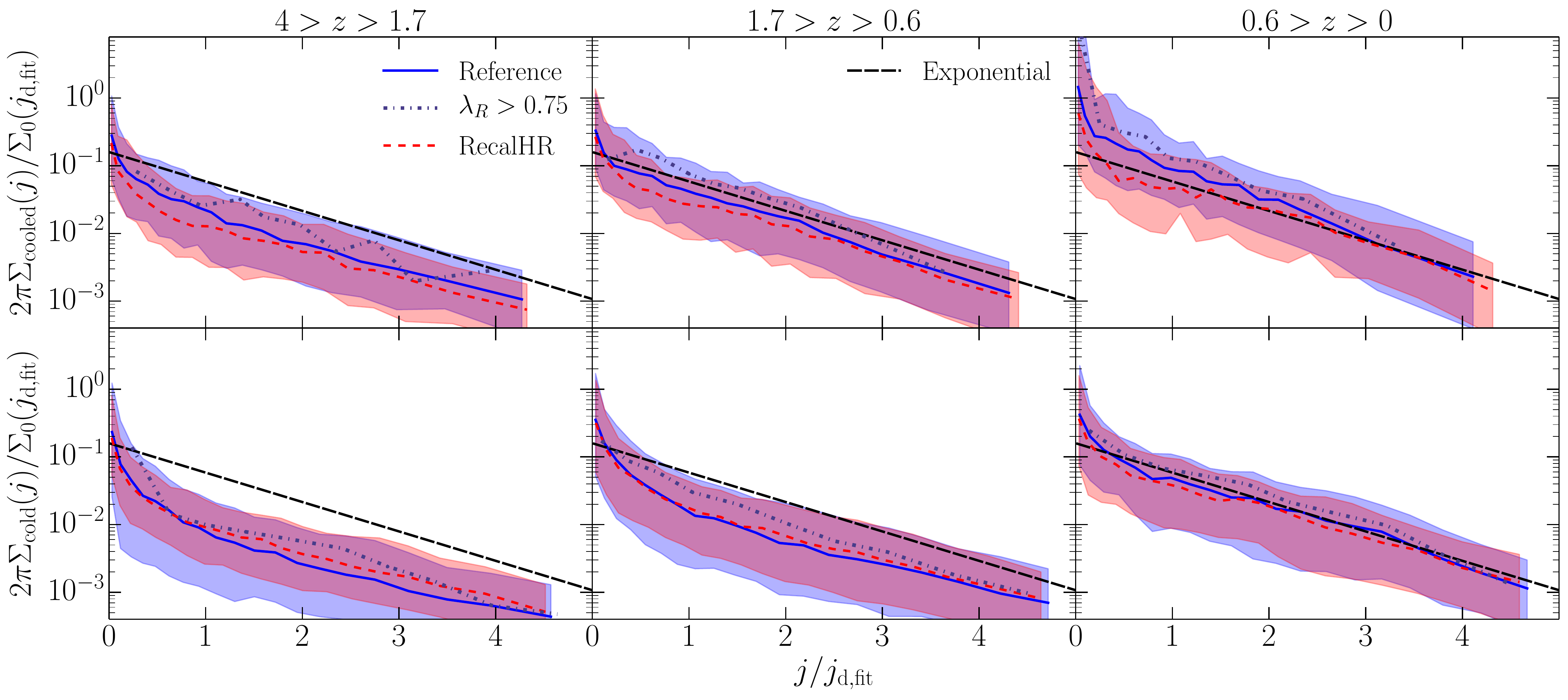}
	\caption{As for Fig.~\ref{fig:Sigma_r}, but now presenting normalised surface density profiles for recently cooled (top panels) and all cold (bottom panels) gas as a function of specific angular momentum.  Because the axes are normalised by best exponential fits, if these profiles were generally well described by an exponential, we should see a tight relation about the long-dashed line in each panel.  Instead, we find an exponential as a function of specific angular momentum is a less accurate description of the profiles to one as a function of radius (cf.~Fig.~\ref{fig:Sigma_r}).}
	\label{fig:Sigma_j}
\end{figure*}

We measure the mean $j$ of particles in each annulus for which we have measured surface density values for our EAGLE galaxies.  This gives us $\Sigma_{\rm cooled}(j)$ and $\Sigma_{\rm cold}(j)$ profiles, to which we fit equation (\ref{eq:Sigma_j}), then normalise these profiles based on the fitted $j_{\rm d}$.  These are presented in Fig.~\ref{fig:Sigma_j}.  At intermediate to low redshifts, an exponential for $\Sigma_{\rm cooled}(j)$ describes the profiles reasonably well, but this is not the case at high redshift.  For $1.7>z>0.6$, there is less of a featured cusp for $\Sigma_{\rm cooled}(j)$ than for $\Sigma_{\rm cooled}(r)$.  This could be explained by the rotational velocities (and hence specific angular momenta) decreasing rapidly at the centres of discs, where pressure support becomes comparable to rotational support.  In general, we find the scatter in the profiles is reduced when considering surface density as a function of radius instead (cf.~Figs \ref{fig:Sigma_r} and \ref{fig:Sigma_j}).  We find these conclusions extend to the overall cold gas profile, and that they apply to the general galaxy population (in our sample) as well as the most rotationally supported systems ($\lambda_R>0.75$, which should be the most discy -- cf.~solid and dot-dashed curves in Fig.~\ref{fig:Sigma_j}).

While our findings suggest one would be better off using equation (\ref{eq:Sigma_cool}) rather than (\ref{eq:Sigma_j}) in a model of galaxy formation, it has been shown that using the latter expression for cooling in a semi-analytic model with disc structure can successfully reproduce the properties of local galaxies, including the relation between total mass and specific angular momentum of stellar discs \citep{stevens16}.  Of course, disc instabilities and feedback will regulate the structure of a galaxy (both as a function of specific angular momentum and radius), so the manner in which gas cools does not provide a complete picture by itself.  An assessment of the importance of details surrounding cooling versus internal regulatory processes in accurately describing galaxy evolution with respect to observations would serve as a useful follow-up study to this work.

\section{Summary}
\label{sec:summary}
We have studied the hot mode of accretion of gas onto galaxies in the EAGLE simulations.  As detailed in Section \ref{ssec:sample}, we selected systems where a sufficient number of gas particles cooled between snapshots in order to learn about the state of haloes undergoing cooling and the galaxies they host.  Our findings presented in this paper can be summarised as follows.

\begin{itemize}

\item The temperature of hot gas in haloes is within a factor of 2 of the virial temperature across all radii (Section \ref{ssec:temphot}).  The gradient of the temperature profiles steepens with time, due in part to stellar feedback.  As centres become hotter, they become less dense, leading to shallower density gradients and the formation of a core in the hot gas (Section \ref{ssec:rhohot}).  The hot-gas density profiles are well approximated by a $\beta$ profile.  We have parametrized this as a simple function of redshift (equation \ref{eq:cfit}), which can easily be adopted by semi-analytic models.

\item Using the precise density, temperature, and metallicity profiles of hot gas in the EAGLE haloes leads to different (and more varied) cooling time profiles than one would obtain assuming a singular isothermal sphere or $\beta$ profile for a halo with the same total mass content, net metallicity, and virial temperature.  At low redshift, these profiles are shallower, leading to cooling radii that are systematically smaller (and, again, more varied), which would translate into lower cooling rates under the model of \citet{white91}.

\item The density profiles of cooling gas remain nearly unchanged with time, irrespective of the underlying hot gas (Section \ref{ssec:temphot}).  While perhaps most alike an exponential, the density profiles of cooling gas are consistent with a singular isothermal sphere (Section \ref{ssec:rhohot}).  So long as models only require cooling gas to look like it is coming from an isothermal sphere, this seems to be a reasonable approximation.

\item Outflows from AGN feedback appear to help regulate the temperature profiles of cooling gas (Section \ref{ssec:temphot}). These become lower and shallower in the centres of haloes over time in response to the decrease in hot-gas density there, which minimises opportunity for cooling.

\item Over the course of cooling, gas loses approximately 60\% of its specific angular momentum, on average (Section \ref{ssec:cons}).  This value is comparable to losses quoted during reports of the `angular momentum catastrophe', yet the galaxies in EAGLE are far from catastrophic.  We find a weak tendency for gas cooling onto rotationally supported galaxies to lose a lesser fraction of its specific angular momentum.  

\item Gas in the process of cooling typically begins with specific angular momentum several times greater than that of the halo.  Its magnitude is largely consistent with the hot gas as a whole, but the spin direction is offset from the rest of the hot gas by tens of degrees (Section \ref{ssec:relj}).  Interestingly, the inner hot gas of the halo is even less well aligned with the cooling gas, despite the cooling gas predominantly originating from the inner halo (cf.~Sections \ref{ssec:rhohot} and \ref{ssec:relj}).  As gas cools, it precesses to become well aligned with the pre-existing gas disc.

\item Recently cooled gas is well approximated by an exponential surface density profile as a function of radius (Section \ref{ssec:Sigma_r}), with the exception of a central cusp, which contains $\sim$20\% of the total cooled mass.  In general, this is more precise than an exponential as a function of specific angular momentum (Section \ref{ssec:Sigma_j}).  Because the cooled gas tends to have higher specific angular momentum than that of the halo, the best-fitting scale radius for these surface density profiles is typically larger than expected from the standard model of disc formation \citep{fall80,mo98}.  This radius is still strongly tied to the spin parameter of a halo, for which we provide a new expression (cf.~equations \ref{eq:rdnew} \& \ref{eq:rdnew_ref}).

\end{itemize}

Our results suggest some of the assumptions surrounding cooling of hot gas in (semi-)analytic galaxy formation models should be revised.  While the parameters of models are often tuned to match the properties of galaxies in the local Universe, by altering the prescriptions surrounding cooling, the star formation histories and higher-redshift properties of these galaxies will change.  This could, therefore, be fruitful in the quest for developing a model of galaxy evolution that can simultaneously explain the high- and low-redshift Universe.

\section*{Acknowledgements}
We thank the city of Perth for its record heat-wave in February 2016 inspiring the title of this paper.  We thank the referee for a constructive report which lead to several improvements in this paper.

This work was supported by a Research Collaboration Award at the University of Western Australia.  Parts of this research were conducted by the Australian Research Council Centre of Excellence for All-sky Astrophysics (CAASTRO), through project number CE110001020.  SC acknowledges support from the CONICYT Doctoral Fellowship Programme.

We acknowledge the Virgo Consortium for making their simulation data available.
We thank PRACE for the access to the Curie facility in France.  
We have used the DiRAC system which is a part of National
E-Infrastructure at Durham University, operated by the Institute
for Computational Cosmology on behalf of the STFC
DiRAC HPC Facility (\url{www.dirac.ac.uk}); the equipment
was funded by BIS National E-infrastructure capital grant
ST/K00042X/1, STFC capital grant ST/H008519/1, STFC
DiRAC Operations grant ST/K003267/1 and Durham University.
The study was sponsored by the Dutch National
Computing Facilities Foundation (NCF) for the use of supercomputer
facilities, with financial support from the Netherlands
Organisation for Scientific Research (NWO), and the
European Research Council under the European Union's
Seventh Framework Programme (FP7/2007- 2013) / ERC
Grant agreement 278594-GasAroundGalaxies.

\end{document}